\PassOptionsToPackage{table,xcdraw}{xcolor} 
\documentclass[acmsmall]{acmart}
\usepackage{multirow}
\usepackage{enumitem}
\usepackage{makecell}

\usepackage{framed}
\definecolor{shadecolor}{RGB}{240, 240, 240}
\definecolor{bordercolor}{RGB}{0, 0, 0}   
\usepackage{CJKutf8}
\renewenvironment{shaded}{%
  \MakeFramed {\advance\hsize-\width \FrameRestore}}%
 {\endMakeFramed}

\newcolumntype{L}[1]{>{\raggedright\arraybackslash}m{#1}}
\newcolumntype{T}[1]{>{\raggedright\arraybackslash}p{#1}}

\definecolor{yellow-gold}{RGB}{217, 164, 78}
\definecolor{olivegreen}{RGB}{141, 163, 104}
\definecolor{redclay}{RGB}{189, 128, 117}

\AtBeginDocument{%

  }

\setcopyright{cc}
\setcctype{by}
\acmJournal{TOCHI}
\acmYear{2026} \acmVolume{1} \acmNumber{1} \acmArticle{1}
\acmMonth{1} \acmPrice{} 
\acmDOI{10.1145/3796234}

\begin{document}

\title{Constella: Supporting Storywriters’ Interconnected Character Creation through LLM-based Multi-Agents}

\author{Syemin Park}
\email{syeminpark@kaist.ac.kr}
\affiliation{%
  \institution{Department of Industrial Design, KAIST}
  \city{Daejeon}
  \country{Republic of Korea}
}

\author{Soobin Park}
\email{soobinpark@kaist.ac.kr}
\affiliation{%
  \institution{Department of Industrial Design, KAIST}
  \city{Daejeon}
  \country{Republic of Korea}
}

\author{Youn-kyung Lim}
\email{younlim@kaist.ac.kr}
\affiliation{%
  \institution{Department of Industrial Design, KAIST}
  \city{Daejeon}
  \country{Republic of Korea}
}

\renewcommand{\shortauthors}{Park et al.}

\begin{abstract}

Creating a cast of characters by attending to their relational dynamics is a critical aspect of most long-form storywriting. However, our formative study (N=14) reveals that writers struggle to envision new characters that could influence existing ones, balance similarities and differences among characters, and intricately flesh out their relationships. Based on these observations, we designed \textit{Constella}, an LLM-based multi-agent tool that supports storywriters’ interconnected character creation process. Constella suggests related characters (FRIENDS DISCOVERY feature), reveals the inner mindscapes of several characters simultaneously (JOURNALS feature), and manifests relationships through inter-character responses (COMMENTS feature). Our 7–8 day deployment study with storywriters (N=11) shows that Constella enabled the creation of expansive communities composed of related characters, facilitated the comparison of characters’ thoughts and emotions, and deepened writers’ understanding of character relationships. We conclude by discussing how multi-agent interactions can help distribute writers’ attention and effort across the character cast.
\end{abstract}



\received{11 June 2025}
\received[revised]{18 December 2025}
\received[accepted]{26 January 2026}

\maketitle


\section{Introduction}
\label{sec:1}
\begin{quote}
\textit{"The single biggest mistake writers make when creating characters is that they think of the hero and all other characters as separate individuals. [...] To create great characters, think of all your characters as part of a web in which each helps define the others."} — John Truby, 2008 \cite{truby2008anatomy}
\end{quote}

Characters are essential building blocks of fictional stories. While the heart of the story may rest in one central character, its pulse comes to life through many. A well-constructed ensemble not only propels the plot \cite{myers2022protagonist} but also creates emotional depth and thematic resonance by allowing characters to challenge, support, and transform one another \cite{mckee2021character, truby2008anatomy, egri1972art, myers2022protagonist}. Thus, crafting an interwoven cast of characters is an important activity for all forms of storytelling—novels, plays, comics, films, and TV shows—but is especially crucial for full-length stories. This is because full-length works that are \textit{"the normal length, rather than being shorter than normal"} \cite{fulllength2025}, provide greater scope for developing the lives of their characters \cite{mckee2021character}. While this process is expressed through terms such as character orchestration \cite{egri1972art}, interconnected web \cite{truby2008anatomy}, and cast formation \cite{mckee2021character}, we refer to it as \textit{interconnected character creation} in this research.

Acknowledging the importance of character creation, Human-Computer Interaction (HCI) researchers have developed computational tools to aid creative writers. One line of work has implemented chatbots embodying writer-defined personas, enabling users to iteratively shape and interact with their characters through free-form conversation \cite{10.1145/3450741.3465253, 10.1145/3613904.3642105, 10.1145/3706598.3713146}. While these works support some relational context (e.g., roleplaying), they primarily focus on single-character development. In contrast, we aim to support character cast creation, and to do so, we see an opportunity for leveraging Large Language Model-based Multi-Agents (LLM-MA). 

LLM-MA are distinguished by their focus on a variety of agent profiles, interactions between agents, and their collective decision-making processes \cite{10.24963/ijcai.2024/890}. Prior HCI research has leveraged LLM-MA to provide users with opinions from diverse viewpoints, finding applications in domains that require a multitude of expert suggestions \cite{park2023choicemates}, audiences \cite{10.1145/3586182.3625114, 10.1145/3706598.3714034}, or simulated users within social computing systems \cite{10.1145/3526113.3545616, 10.1145/3613904.3642545, 10.1145/3586183.3606763}. Although these works are not directly related to the storytelling context, they collectively underscore the potential of LLM-MA when used for interconnected character creation. As noted by storywriting literature, \textit{"the most important step in creating your hero, as well as all other characters, is to connect and compare each to the others"} \cite{truby2008anatomy}. Thus, by enabling the simultaneous exploration of contrasting character perspectives, LLM-MA can empower writers to construct a complex, multilayered cast of characters. 

To design an LLM-MA tool that supports storywriters’ interconnected character creation, we first sought to understand the challenges they face during this process and ground our tool in the strategies they find effective for overcoming these difficulties. To this end, we conducted a formative study with 14 writers. We found that writers address key challenges through the following strategies: 1) To conceptualize new characters capable of influencing existing ones, they brainstorm a wide range of related characters. 2) To balance similarities and differences between characters, they imagine each character’s inner world and how each would respond uniquely to the same situation. 3) To flesh out character relationships, they often create moments of interaction regardless of the main plot to gain insight into their dynamics.

With these observations, we derived three design goals for an LLM-MA tool to support interconnected character creation: 1) Suggest diverse characters related to existing ones, 2) support the comparison and investigation of in-depth character identities, and 3) manifest character relationships through exploratory interactions. To translate these goals into concrete system features, we drew inspiration from interaction concepts found in social media (e.g., 
\cite{facebook2025, threads2025, reddit2025}), as they provide effective means for designing social interactions among story characters that align with each design goal. Specifically, these platforms enable users to connect with individuals beyond their usual acquaintances through potential connection points, freely express differing or similar thoughts on topics of mutual interest, and display or reinforce relationships by reacting to others’ content. We anticipated that drawing on these affordances to simulate character interactions would intuitively facilitate our design goals by echoing everyday social behavior.

\begin{figure}
  \includegraphics[width=\textwidth]{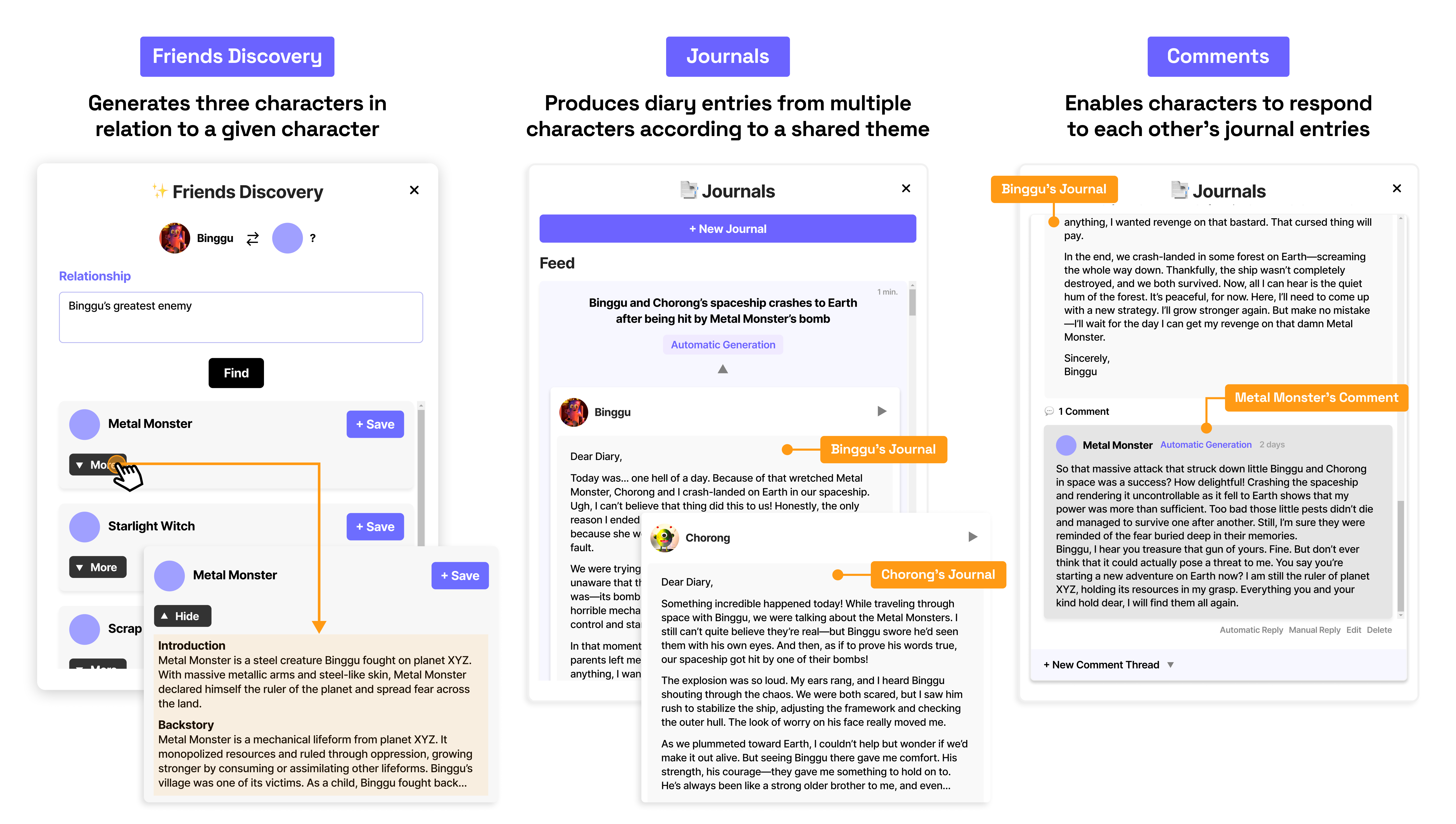}
  \caption{Three main features of Constella that support storywriters’ interconnected character creation. (Left) FRIENDS DISCOVERY generates three characters in relation to a given character, facilitating character cast expansion. (Center) JOURNALS produces diary entries from multiple characters according to a shared theme, allowing the comparison of their inner mindscapes. (Right) COMMENTS enables characters to respond to each other’s journal entries, surfacing deeper layers of character relationships.}
  \Description{The Friends Discovery section shows new characters (Metal Monster, Starlight Witch, and Scrap Metal Prince) generated in relation to Binggu. The Journals section shows two journal entries, one written by Binggu and one written by Chorong, with the theme set to ‘Binggu and Chorong’s spaceship crashes to Earth after being hit by Metal Monster’s bomb.’ The Comments section shows Metal Monster’s response to Binggu’s journal entry.}
  \label{fig:teaser}
\end{figure}

Thus, we propose \textit{Constella}\footnote{Constella is short for constellation, underscoring that characters, like stars, gain new meaning as they connect, relate, and are viewed together.}, an LLM-MA tool that supports storywriters’ interconnected character creation through social media-inspired interactions (\autoref{fig:teaser}). Constella enables writers to iteratively build characters and their relationships via dedicated character profiles. Upon creating a profile, the FRIENDS DISCOVERY feature helps writers uncover new characters linked to existing ones by generating their mini profiles. Writers can also use the JOURNALS feature to contrast the inner mindscapes of characters by generating their diary entries. Lastly, writers can elaborate on character relationships as they respond to each other’s journal entries through the COMMENTS feature. Together, these features offer resources for writers to flexibly expand, deeply inspect, and tangibly concretize their character casts throughout their development process.

To explore how our tool integrates into storywriters’ practices, we conducted a 7–8 day deployment study with 11 writers. During the study, participants used the tool through tasks that included creating multiple characters, a backstory, scenes, and a story outline, all within the context of writing a full-length fictional story. The results show that Constella supported diverse approaches to interconnected character creation. It facilitated the construction of multiple worlds populated by distinct yet interlinked characters, supported the analysis of character differences and similarities that inspired new scenes, and enabled deeper exploration of character relationships by surfacing emotional tensions. Notably, our tool encouraged writers to focus on a broader range of characters, including those that might otherwise have remained underdeveloped. Constella was also perceived to play a complementary role in the co-creative process without compromising their sense of authorial agency. We conclude by discussing the potential of LLM-MA to help writers distribute creative attention across the entire character cast and how AI-infused writing tools can be designed to support authorial agency. 

This paper presents the following contributions:

\setlist[enumerate,1]{      
  label=(\arabic*),         
  leftmargin=*,             
  labelsep=.5em,            
  itemsep=0pt,              
  topsep=0pt                
}
\begin{enumerate}
  \item Constella, an LLM-MA tool that assists storywriters in constructing character casts through LLM-MA.

  \item Empirical insights from a deployment study illustrating Constella’s integration with storywriters’ practices. 

  \item Design implications for AI-based writing support tools to help preserve authorial agency.
\end{enumerate}

\section{Background and Related Work}
\label{sec:2}
This work aims to support storywriters’ interconnected character creation process through LLM-MA. In this section, we first outline the theoretical foundations of interconnected character creation in storywriting (§\ref{sec:2.1}). We then review prior AI tools designed to assist character creation (§\ref{sec:2.2}). Finally, we position our work within the broader landscape of AI-assisted writing tools that move beyond ready-made text generation to support creative processes (§\ref{sec:2.3}).

\subsection{Interconnected Character Creation}
\label{sec:2.1}

Narratives consist of actions, happenings, characters, settings, and discourse \cite{chatman1980story}. Our research focuses on the element of characters, which can be defined through archetypes\footnote{An archetype is \textit{"a primordial image, character, or pattern of circumstances that recurs throughout literature and thought consistently enough to be considered a universal concept or situation."} \cite{britannicaArchetype}.} \cite{lee2013psychology, mckee2021character, truby2008anatomy, myers2022protagonist, schmidt201245}, backstories \cite{lee2013psychology, mckee2010story, mckee2021character, truby2008anatomy, card2010elements, seger1990creating, corbett2013art, myers2022protagonist}, and various physical, psychological, and sociological attributes \cite{chatman1980story, mckee2010story, card2010elements, seger1990creating, egri1972art, field2005screenplay, barker2019build, edelstein2006writer}.

As full-length fictional stories typically feature an ensemble of characters, writing guides offer suggestions for developing characters in relation to one another \cite{mckee2021character, truby2008anatomy, egri1972art, myers2022protagonist, eastman2021shakespeare}. For instance, McKee \cite{mckee2010story} uses the metaphor of a solar system, suggesting that writers imagine the main character as a star and other characters as orbiting planets, each positioned based on their influence on the protagonist and one another. He advises writers to begin by creating the central role, followed by the incremental development of supporting roles. In contrast, Truby \cite{truby2008anatomy} stresses the importance of developing characters simultaneously, encouraging writers to draw continuous comparisons and connections among them. He argues that this process highlights each character’s individuality, allowing even supporting characters to become more distinct and meaningful alongside the protagonist.

While literature on storywriting underscores the value of creating characters in an interconnected manner, it tends to focus on general principles rather than providing comprehensive explanations of the real-world challenges writers face during interconnected character creation and the underlying causes of these difficulties. Furthermore, the degree to which these theoretical approaches reflect actual writing practices remains uncertain. As prior HCI research has highlighted the diversity of writers’ practices and the variability of their needs \cite{ippolito2022creative, 10.1145/3698061.3726910, 10.1145/3544548.3580782}, we anticipated discrepancies between the formalized processes outlined in theory and the actual practices of writers in the wild. Thus, to design a tool that effectively supports storywriters, we determined it was essential to draw not only on theoretical insights but also to engage deeply with empirical evidence. Based on this background, we grounded the development of our tool through in-depth interviews with storywriters, using the previously cited writing guides as references when developing the interview questions. 

\subsection{AI Support for Character Construction}
\label{sec:2.2}
In this section, we review LLM-based tools designed for storywriters, focusing on systems that support the creation of new characters or facilitate character interactions through agents that embody writer-defined personas. 

Several works have utilized Large Language Models (LLMs) to generate new characters \cite{10.1145/3544548.3581225, 10.1145/3654777.3676352, kreminski2024intent, swanson-etal-2021-story, sudowrite2025, chen-etal-2024-hollmwood, huot2025agents}, often drawing from other elements of the story to align with the writers’ overall vision and storytelling goals. For example, \textit{Dramatron} \cite{10.1145/3544548.3581225} allows users to first define a logline\footnote{A logline is \textit{"a single sentence describing your entire story"} \cite{russell2000screen}, such as what the protagonist wants or desires and what is at stake \cite{bishop2016sell}.}, which is then used to generate not only the characters but also the title and plot. \textit{Patchview} \cite{10.1145/3654777.3676352} enables writers to create concepts, such as good and evil, and choose positions along these spectra, guiding the system to create appropriate worldbuilding elements, including characters. However, existing works do not explicitly focus on user-specified relational ties between characters as a primary axis for character generation, despite the central role they play in character development and plot progression \cite{varotsi2019conceptualisation, truby2008anatomy, myers2022protagonist, eastman2021shakespeare, harvey1970character}.

Another line of work has aimed to enable storywriter-character interactions through chatbots designed to reflect writer-defined attributes \cite{10.1145/3450741.3465253, 10.1145/3613904.3642105, banterbot2019, 10.1145/3532106.3533506, stark2023can}. Most relatedly, \textit{CharacterChat} \cite{10.1145/3450741.3465253} first introduced the concept of \textit{"progressive manifestation,"} where the writers gradually shape a chatbot into their character by adding new attributes through back-and-forth conversation. \textit{CharacterMeet} \cite{10.1145/3613904.3642105} extends this research by utilizing an LLM and implementing a conversation context, enabling writers to roleplay as other characters. However, with the exception of those created for children \cite{10.1145/3706599.3719932}, implementations of agent-based character embodiments for storywriters \cite{10.1145/3450741.3465253, 10.1145/3613904.3642105, banterbot2019} have mostly concentrated on single-character development. 

While we focus on providing support for fictional storywriters, prior research in the domain of interactive storytelling has proposed tools \cite{10.1145/3649921.3656987, sun2025drama, 10.1145/3706598.3713363, han-etal-2024-ibsen, 10.1145/3746058.3758995} that utilize multiple character embodiments via LLM-MA. As the emergent nature of LLM-MA poses challenges for writers to steer the direction of interactive stories (e.g., games), these works focus on providing control mechanisms that mediate between the author’s intent, the unpredictability of LLM-based agents, and player-introduced actions. For example, \textit{Storyverse} \cite{10.1145/3649921.3656987} translates high-level plot outlines into characters’ action sequences during playtime, while the game player can also steer the unfolding events, such as by killing a character. 

Beyond interactive storytelling, several works \cite{yu-etal-2025-multi, chen-etal-2024-hollmwood, huot2025agents} adapt LLM-MA for fictional storywriting, but their emphasis is on proposing automated story generation pipelines, thereby providing limited support for writers’ iterative and incremental character construction processes \cite{kreminski-martens-2022-unmet}. For instance, Yu et al.’s work \cite{yu-etal-2025-multi} assumes that detailed character personas and the plot outline are readily available and proceeds to generate inter-character simulations that produce the detailed text for the story. Other systems \cite{chen-etal-2024-hollmwood, huot2025agents} limit the writer’s contribution to supplying a seed storyline; from that point onward, agents automatically detail the plot and characters, multi-agent exchanges generate the dialogue, and a final agent combines them into the complete narrative. 

While these approaches demonstrate the potential of LLM-MA in storywriting, they primarily utilize LLM-MA simulations as either real-time, emergent narratives or as components within largely automated pipelines for complete story generation. In contrast, we focus on leveraging LLM-MA to support fictional storywriters in constructing their character casts, specifically through the design of scaffolding materials (e.g., journal entries and comments) rather than final story text. This reflects our deliberate intent to uphold authorial agency by preserving the writers’ role in the expressive act of the writing process. In line with this focus, we investigate how our tool supports writers through a deployment study, which, to our knowledge, is the first research to offer empirical insights into how a character-based LLM-MA tool integrates into the real-world practices of fictional storywriters.

\subsection{Assisting Writers Beyond Generating Ready-Made Text}
\label{sec:2.3}
Recent advancements in language models have spurred the development of various tools aimed at assisting writers. One predominant trend in this space involves co-writing systems that offer suggestions in the form of output-ready text, such as next sentence generation \cite{swanson-etal-2021-story, 10.1145/3491102.3502030, 10.1145/3490099.3511105, 10.1145/3172944.3172983, yang2022ai, 10.1145/3180308.3180329, calderwood2020novelists, 10.1145/3462204.3481771}, text variation \cite{swanson-etal-2021-story, 10.1145/3490099.3511105, 10.1145/2757226.2757229, 10.1145/3613904.3641899}, and narrative generation \cite{10.1145/3544548.3581225, 10.1145/3649921.3656987, 10.1007/978-3-031-76806-4_7, sun2025drama, 10.1145/3706598.3713363, 10.1145/3411763.3450391, yu-etal-2025-multi, chen-etal-2024-hollmwood, huot2025agents, han-etal-2024-ibsen}. However, recent studies investigating writers’ perceptions of AI systems have highlighted the importance of the \textit{"emotional values of writing"} \cite{10.1145/3532106.3533506}, values that may be diminished when AI systems reduce opportunities for writers to \textit{"express their ideas on their own"} \cite{10.1145/3532106.3533506}. 

In response, a growing body of research \cite{stark2023can, 10.1145/3635636.3656187, arnold2021generative, 10.1145/3526113.3545672, kim2024towards, 10.1145/3698061.3726916, 10.1145/3706598.3713161} has been actively proposing alternative ways for generative models to support writers without direct text generation; that is, text that can be immediately inserted into their final work without further modification. For example, AI tools that provide experience advising \cite{stark2023can}, questions \cite{arnold2021generative}, or continuous summaries to support process-oriented reflection \cite{10.1145/3526113.3545672} have been introduced. Collectively, these approaches are promising ways to help writers retain a strong sense of control over their work.

We situate our tool in this body of work by proposing a Creativity Support Tool (CST) \cite{10.1145/1323688.1323689, 10.1145/3290605.3300619, 10.1145/3461778.3462050} that offers open-ended resources for storywriters. Specifically, we provide resources in social media-inspired formats such as mini profiles, diaries, and comments, which likely encourage writers to iterate over, adapt, and interpret the generated content rather than directly copying and inserting it as-is. This offers a way to support a form of agency that has been found to be vital in AI-assisted storywriting yet relatively underexplored in prior HCI research: writing process control \cite{carrera2025add}, which is maintained when writers treat generated content as raw material to be further shaped by themselves. In doing so, our approach makes room for the author’s expressive intent to flow through the interpretation of underdetermined outcomes \cite{10.1145/3690712.3690725}.

\section{Formative Study}
\label{sec:3}
To inform the design of an LLM-MA tool that supports interconnected character creation, we conducted semi-structured interviews with storywriters. We focused on identifying their approaches to interconnected character creation, the challenges they face during this process, and the strategies they use to overcome these difficulties. Insights from this study informed the three design goals for our tool. The study process was approved by our university’s Institutional Review Board (IRB No. KH2024-115).

\subsection{Participants}
\label{sec:3.1}
We recruited writers with prior experience in creating full-length fictional stories. Considering the domain of storytelling, we aimed to gather insights across diverse story types and genres. Furthermore, we aimed to diversify participants based on their total duration of creative writing experience to gain a comprehensive understanding from individuals relatively newer to storywriting to those with greater familiarity. Lastly, we recruited participants with and without formal creative writing education, as well as those with and without publication experience. This inclusive approach was informed by Gero et al.’s work \cite{10.1145/3544548.3580782}, which argues that formal education does not necessarily equate to expertise and that relying on publication experience as a metric for expertise could unnecessarily marginalize writers.
Participants were recruited through online writer communities, university communities, personal contacts, and social media posts uploaded by researchers and their acquaintances. All participants were selected through purposeful sampling \cite{palinkas2015purposeful} to ensure maximum variation across story types, genres, and total duration of creative writing experience.

In total, 14 participants (F=8, M=6) were recruited (\autoref{table:formative-demo}). Story types ranged from films (N=11), TV shows (N=10), novels (N=10), plays (N=3), comics (N=2), and animation (N=1). The total number of writing genres they had experience with reached 32, including examples such as comedy, romance, and historical fiction. The duration of their creative writing experience ranged from 2 to 26 years, with an average of 11.93 years. All participants resided in South Korea. 

\begin{table}[ht]
    \renewcommand{\arraystretch}{1.4}
    \caption{Formative study participants’ self-reported demographics.}
    \label{table:formative-demo}
    \resizebox{\textwidth}{!}{%
    \begin{tabular}{@{}lll m{3.5cm} m{5.5cm} m{3.1cm}
                    m{1.8cm} m{1.8cm}@{}}
    \toprule
    \textbf{ID} & \textbf{Gender} & \textbf{Age} & \textbf{Story Types} & \textbf{Genres} &
    \raggedright \textbf{Years of Creative}\\[-0.5ex]\raggedright \textbf{Writing Experience} &
    \raggedright \textbf{Related}\\[-0.5ex]\raggedright \textbf{Degree} &
    \textbf{Publishing Experience} \\
    \midrule
    P1  & Male   & 41 & TV Show, Film, Novel &
          Drama, Melodrama, Comedy, Action Noir & 21 & Bachelor’s & Yes \\
    P2  & Female & 41 & Film, TV Show, Novel &
          Romantic Comedy, Melodrama, Science Fiction, Humanist Drama & 21 & Bachelor’s & Yes \\
    P3  & Male   & 27 & Film &
          TV Show, Thriller, Horror, Romance, Experimental Film & 10 & Master’s & Yes \\
    P4  & Male   & 51 & TV Show, Film &
          Horror, Historical Fiction, Courtroom Drama, Comedy, Heist Film, Romantic Comedy & 21 & Master’s & Yes \\
    P5  & Female & 25 & Novel, Film &
          Thriller, Mystery, Comedy, Romance & 7 & Bachelor’s & Yes \\
    P6  & Male   & 46 & Comics, TV Show, Novel &
          Action, Noir & 26 & Bachelor’s & Yes \\
    P7  & Female & 51 & Animation, TV Show, Novel &
          Historical Fiction, Science Fiction, Fantasy, Coming-of-age Drama, Horror, Thriller, Slice-of-life Mystery, Educational Comics, Fantasy Adventure, Revenge Drama, Social Drama, Office Drama, Romance & 15 & None & Yes \\
    P8  & Female & 31 & Play, Film, TV Show, Novel &
          Drama & 10 & Master’s & Yes \\
    P9  & Female & 42 & Play, TV Show, Film &
          Humanist Drama, Coming-of-age Drama, Family Drama, Historical Fiction, Thriller, Occult, Action & 14 & None & Yes \\
    P10 & Female   & 46 & TV Show, Film &
          Romance, Horror, Mystery, School Drama & 6 & None & Yes \\
    P11 & Male   & 53 & Novel, Film, TV Show &
          Thriller, Coming-of-age Drama, Comedy, Action & 2 & None & No \\
    P12 & Male   & 36 & Novel &
          Literary Fiction, Fantasy & 2 & None & No \\
    P13 & Female & 26 & Comics, Novel, Play, Film &
          Drama, Science Fiction, Romance Thriller & 7 & Bachelor’s & No \\
    P14 & Female & 26 & Novel, Film, TV Show &
          Drama, Social Problem Film, Coming-of-age Drama, Melodrama, Comedy, Art Film & 5 & Bachelor’s & Yes \\
    \bottomrule
    \end{tabular}}%
\end{table}

\clearpage

\subsection{Procedure}
\label{sec:3.2}
\begin{figure}[H]
    \centering
    \includegraphics[width=\textwidth]{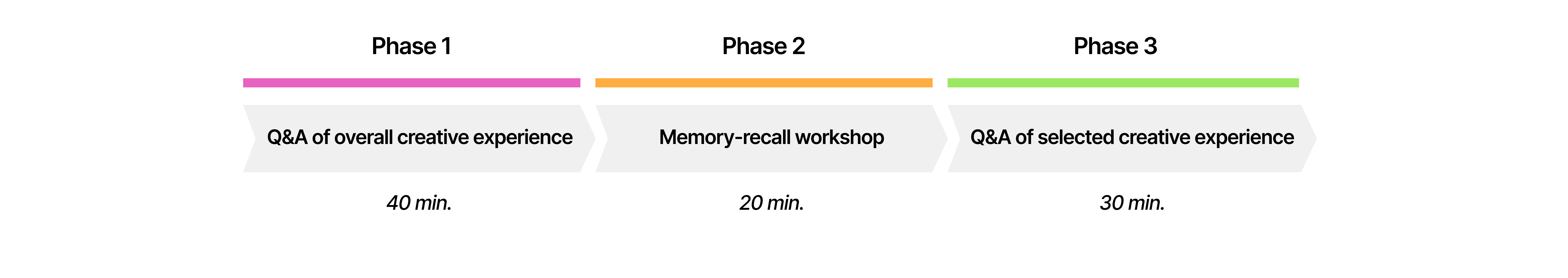}
    \caption{Procedure of the formative study.}
    \Description{Fully described in the text.}
    \label{fig:formative-process}
\end{figure}

The formative study was divided into three phases: (1) Q\&A of overall creative experience, (2) memory-recall workshop, and (3) Q\&A of selected creative experience (\autoref{fig:formative-process}). All three phases were conducted sequentially within one day. Each participant completed the study either in person (N=5) or remotely via Zoom (N=9), depending on participants’ preferences.

\textbf{(1) Q\&A of overall creative experience (40 minutes)}: After obtaining informed consent, we introduced the goal of the study. To gain overarching insights into their interconnected character creation process, we asked participants to answer the questions (\autoref{table:formative-interview} in Appendix \ref{sec:Appendix-questions}) based on all their full-length stories. The questions inquired about their experiences in developing characters with differing roles, comparing characters, and building character relationships. 

\begin{figure} [H]
    \centering
    \includegraphics[width=\textwidth]{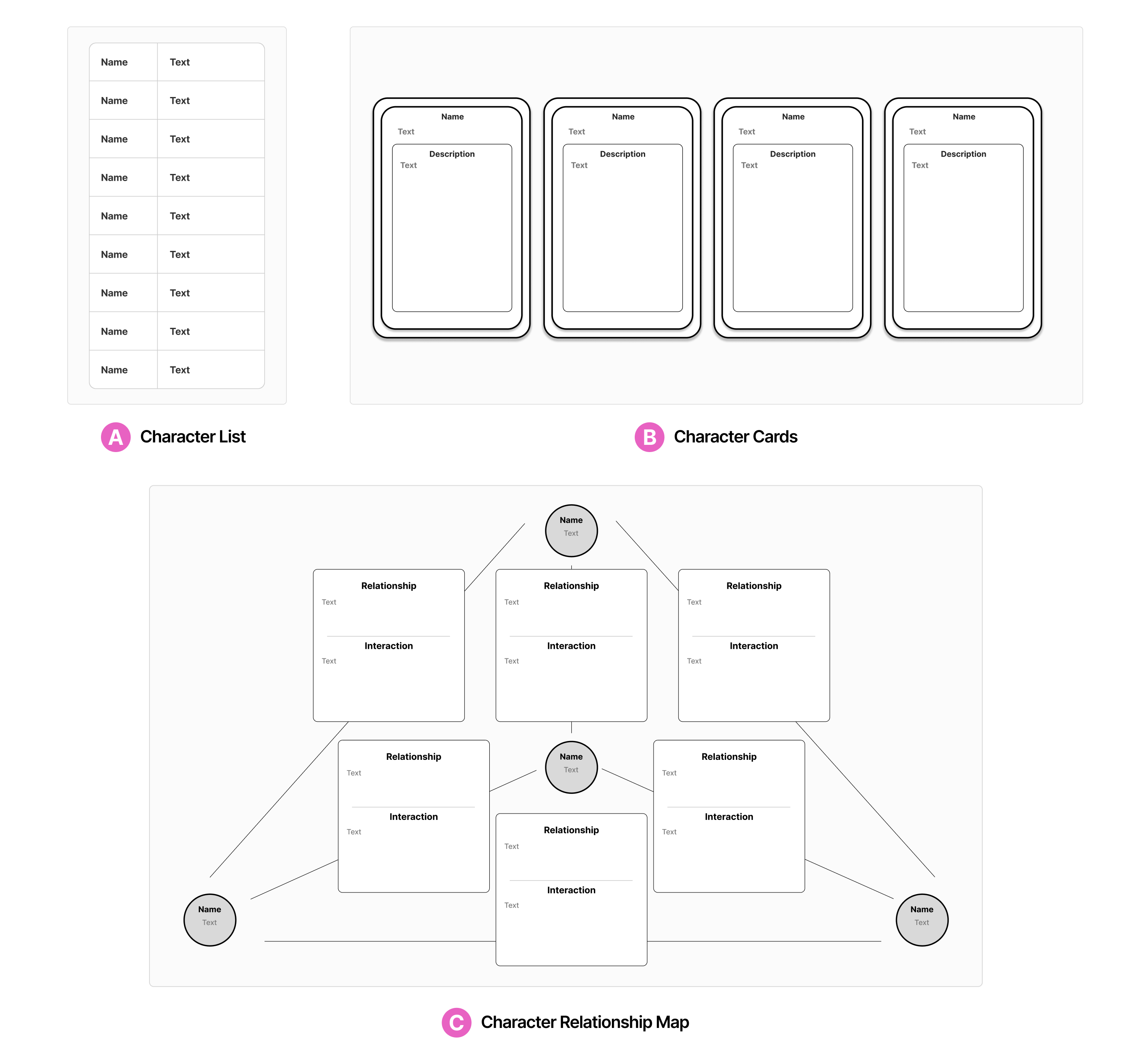}
    \caption{Materials for the workshop. (A) Character List. (B) Character Cards. (C) Character Relationship Map.}
    \Description{The Character List is a table with an editable Name column. Character cards include fields for a character’s name and description. The character relationship map includes fields to enter multiple character names and specify their relationships and interactions.}
    \label{fig:formative-materials}
\end{figure}

\textbf{(2) Memory-recall workshop (20 minutes)}: To help participants recall their work in preparation for the next phase, we offered worksheets that served as elicitation materials. Participants selected one of their previously created full-length stories and completed three types of worksheets: the Character List, Character Cards, and Character Relationship Map (\autoref{fig:formative-materials}). For online participants, the worksheets were shared via FigJam \cite{figjam2025}, while offline participants received printed copies. Further details of the workshop materials are described in §\ref{sec:3.3}.

\textbf{(3) Q\&A of selected creative experience (30 minutes)}: In this phase, we aimed to gain a deeper understanding of their interconnected character creation process through concrete, situated examples. Thus, participants answered the same set of questions (\autoref{table:formative-interview} in Appendix \ref{sec:Appendix-questions}), but now contextualized to the specific story they had selected in the second phase. All participants received compensation of 30,000 KRW (approx. 21 USD) for their participation.

\subsection{Materials for Memory-Recall Workshop}
\label{sec:3.3}

The worksheets for the memory-recall workshop were designed to help participants reflect on the characters from one of their previously created full-length stories. The \textbf{Character List} helped participants recall both main and supporting characters by listing their names. The \textbf{Character Cards} prompted focused reflection on 3–4 main characters by writing their descriptions, an approach inspired by character sheets used by storywriters \cite{10.1145/3706598.3713161}. The \textbf{Character Relationship Map} guided participants in outlining relationships and interactions among the main characters, inspired by character maps discussed in storywriting literature \cite{myers2022protagonist}. Acknowledging that these materials may not align with their usual practices, participants were encouraged to recall their work and creation process using alternative methods if preferred.

\subsection{Analysis}
\label{sec:3.4}
We transcribed over 1,260 minutes of audio recordings from interviews with 14 participants. The worksheets completed during the memory-recall workshop were used solely as reference materials to clarify participants’ responses during the transcription process.

The interview data were qualitatively analyzed through three cycles of coding. In the first cycle, the first author used structural coding \cite{saldana2021coding} to categorize sections of the data in relation to our study’s goal (approaches, challenges, strategies), resulting in our initial codes. Based on these initial codes, the first author developed sub-codes through process coding \cite{miles2013qualitative} to identify patterns in participants’ conceptual and observable actions. These sub-codes were then iteratively refined through discussions among the researchers. Finally, the first author used pattern coding \cite{saldana2021coding} to cluster the sub-codes into three overarching categories that explain participants’ interconnected character creation practices. This final analysis process was also reviewed by the other researchers. The entire analysis process was conducted over two months.

\subsection{Findings}
\label{sec:3.5}
We report participants’ approaches to interconnected character creation, the challenges they face during this process, and the strategies they use to address these difficulties. These insights are organized into three overarching categories that encapsulate their practices.
    \subsubsection{Branching Out Related Characters}
    \label{sec:3.5.1}
    \paragraph*{\textbf{Approach}} In line with writing theory on character cast construction \cite{mckee2021character}, writers would often start by creating their main characters, then branch out new characters in relation to these existing characters (P1, P3, P4, P5, P6, P7, P8, P9, P10, P11). One specific method involved splitting the traits of the main character and distributing them among other characters (P7, P9). P9 mentioned that this method ensured that these characters would form a cohesive whole, as each character would reflect different aspects of the main character’s personality. Another method was to think of new characters that would complement existing characters or have opposing characteristics to create conflict (P1, P5, P6, P7, P9, P10). One last notable method was to imagine specific situations where the main characters would be placed and envision other characters that would naturally exist in the same environment (P1, P7, P9). 

    \paragraph*{\textbf{Challenges}} However, thinking of new characters, especially those who would influence the protagonist in specific ways, was often challenging for some writers (P4, P7). The guidance offered to the main character and the inner conflicts they experience depend on the supporting characters. Therefore, participants placed great importance on finding characters that could appropriately fulfill these roles. 
    
    \begin{quote}
        \textit{"The character needs to act in a certain way, but who could be the essential character to influence that character towards this direction? A friend, perhaps? What kind of personality and values does this friend have, and how would they influence the protagonist? It’s not just about what they say; it’s also about the kind of relationship they’ve had until now. But who could that be?"} (P4)
    \end{quote}
    P7 also shared her experience, mentioning that the process of adjusting supporting characters so they fit these roles seems \textit{"roughly correct at first, but roughly correct is not enough. One must continuously refine these characters and repeat this process."}
    
    \paragraph*{\textbf{Strategies}} In response to this challenge, P4 said he would brainstorm a wide range of characters with a fellow writer, see how each character "feels," and swap them out until something clicks. P7 added that she would sometimes even change the settings of existing characters when things were not working out.

    In summary, writers often expand their character casts in relation to the main characters. However, coming up with new characters that could appropriately influence these core characters was often challenging. This process led to ideating a broad spectrum of potential characters and iteratively refining them to create the perfect match. Thus, the tool should offer a wide selection of related characters with key descriptions, allowing writers to quickly explore options and iteratively adapt characters when needed. 

    \subsubsection{Balancing Similarities and Differences}
    \label{sec:3.5.2}
    \paragraph*{\textbf{Approach}} Many participants explained that they often constructed characters by comparing their identities (P1, P2, P3, P5, P7, P8, P10, P13). This practice serves two main purposes. First, it helps ensure that each character has distinct roles and personalities. For example, P1 would start by thinking of ways his characters would contrast visually and progressively alter several attributes of each character, such as their gender, when their identities overlapped. Second, this approach enables participants to create characters who not only contrast but also share values and beliefs. P10 emphasized the importance of this practice for creating complex characters who both attract and clash with each other. Taken together, these practices mirror the theoretical concepts of \textit{character orchestration} and the \textit{unity of opposites} \cite{egri1972art}, which emphasize arranging contrasting yet interdependent characters to sustain dramatic conflict.

    \paragraph*{\textbf{Challenges}} Despite these initial approaches, many writers found it challenging to create distinct characters, leading to overlapping character identities (P1, P2, P5, P7, P8, P10, P11, P13). For instance, P2 worried that her characters would reflect her own values and thoughts too much, further explaining that individuals created from a single source would inevitably result in similar personalities. In addition, writers explained that balancing character identities that would contrast and align in different aspects was also challenging (P7, P8, P13). 

    \begin{quote}
    \textit{"What feels more compelling to me is exploring the sparks that fly when two equally powerful beings, each acting based on their beliefs or convictions, come into conflict. So, it’s not just about aligning their social status but also about balancing the scale of their beliefs and actions, which is what I find challenging."} (P8) 
    \end{quote}

    \paragraph*{\textbf{Strategies}} To create distinct character identities, writers aimed to capture essential traits that could clearly define their characters through expert interviews, reading other stories, and observing social media, among other sources. In addition, writers (P1, P2, P3, P5, P7, P8, P9, P10) expressed the importance of becoming immersed in the character’s inner world and deeply empathizing with them, such as by writing character backstories in the first person. To create characters that hold both contrasting and shared identities, writers (P1, P2, P3, P4, P7, P8, P9) would place characters in identical situations to further establish their personalities, a tactic also described as \textit{"character webs"} in writing theory \cite{eastman2021shakespeare}. Some writers used predefined questionnaires and answered these questions for each character in the initial stage of characterization. Others would create such situations as scenes within their stories to continuously refine character details.

    In short, writers often compared character identities to create both distinguishable and similar characters. Yet, balancing character similarities and differences remained a persistent challenge. Strategies to tackle this issue included examining how several characters would respond to a common situation, as well as deeply empathizing with and understanding characters’ inner worlds. Therefore, the tool should also allow writers to explore how each character interprets identical contexts in ways that reveal their internal thought processes and subjective perspectives. 

    \subsubsection{Fleshing Out Relationships}
    \label{sec:3.5.3}
    \paragraph*{\textbf{Approach}} Consistent with prior work \cite{10.1145/3706598.3713161} that investigated writers’ use of tools for ideation, we found that several writers created character relationship maps in the form of diagrams to organize and explore new connections between characters (P1, P5, P9, P13, P14). Specifically, P5 would use these diagrams to get a broader view of character relationships and often eliminate characters who remain isolated and unconnected to others. Similarly, P13 and P14 noted that these diagrams help them discover new relationships between characters that they had not previously considered. 

    \paragraph*{\textbf{Challenges}} Despite these approaches, intricately fleshing out character relationships remained a challenge for several writers (P5, P7, P9, P11, P13). For example, P9 explained: 

    \begin{quote}
        \textit{"Characters A and B may be defined as close friends, but within that, there might be nuances, like one of them not attending the other’s wedding. Even though they might all be grouped as close friends from the outside, the deeper you delve, the more you uncover different circumstances and histories that make the relationships more complex."} (P9)
    \end{quote}

    \paragraph*{\textbf{Strategies}} To tackle these challenges, one major strategy used by writers (P9, P13, P14) was to ideate a wide range of character interactions to deepen their understanding of these relationships, even if they did not make it into the final narrative. Similar to the \textit{"free scene"} \cite{myers2022protagonist} method in which writers draft exploratory scenes that place characters in imagined situations to observe how they behave, P13 referred to this technique as \textit{"hypothetical situations"} and employed it to determine whether the main characters would get along before drafting the main plot. In addition, P9 mentioned that the emotions and interactions between characters would slowly accumulate through these activities and naturally evolve into specific dialogue or episodes within the story.

    To summarize, character relationship maps were utilized to identify and organize connections between characters. However, writers struggled to thoroughly build upon these relationships. Therefore, participants imagined how characters would interact with one another in various situations, regardless of the story’s overall flow. This implies that the tool should also aim to reveal character relationships through open-ended reactions towards each other. 

\subsection{Design Goals}
\label{sec:3.6}
Based on these findings, we identified three design goals (DG) to build a tool to support storywriters’ interconnected character creation. \newline
\begin{enumerate}[label=DG~\arabic*.,  
                 labelsep=0.6em,       
                 itemsep=0.8\baselineskip] 
\item \textbf{Suggest diverse characters related to existing ones.}
      To support writers in branching out related characters, the tool should focus on broadly recommending a wide range of related characters with core traits, rather than extensive information about a single character (§\ref{sec:3.5.1}). This goal focuses on the breadth of information.

\item \textbf{Support the comparison and investigation of in-depth character identities.}
      To assist writers in balancing the similarities and differences between characters, the tool should illustrate how each character interprets the same context in ways that reveal their inner mindscapes (§\ref{sec:3.5.2}). This goal emphasizes the depth of information.  

\item \textbf{Manifest character relationships through exploratory interactions.}
 To help writers flesh out character relationships, the tool should show open-ended reactions between characters as testbeds for building and refining their dynamics (§\ref{sec:3.5.3}). This goal emphasizes speculative information rather than narrative progression. \newline
 
\end{enumerate}
Through these three design goals, we aim to provide a plethora of resources that storywriters can mold into concrete forms, use to understand their work deeply, or draw inspiration from while leaving the generated results as-is.

\section{Constella}
\label{sec:4}
Based on the design goals, we developed Constella, a creativity support tool designed to assist storywriters’ interconnected character creation process through LLM-MA. Its core features include FRIENDS DISCOVERY, JOURNALS, and COMMENTS. FRIENDS DISCOVERY generates three characters in relation to a given character, aiding writers in expanding their ensemble with diverse additions (\textbf{DG 1}). JOURNALS produces diary entries from multiple characters according to a shared theme, revealing their internal emotions, thoughts, and behaviors (\textbf{DG 2}). COMMENTS enables characters to respond to other characters’ journal entries, surfacing additional layers of character relationships (\textbf{DG 3}). Here, we first discuss our source of inspiration when translating the design goals into system features (§\ref{sec:4.1}). We then present the system’s components (§\ref{sec:4.2}), an example use scenario (§\ref{sec:4.3}), and the implementation and prompt design (§\ref{sec:4.4}).

\subsection{Metaphor and Interaction Concept}
\label{sec:4.1}
In this section, we explain the rationale for adopting social media as an overarching metaphor to address our design goals. We also illustrate how specific social media interaction concepts inspired the main features of our LLM-MA tool.

Metaphors have been utilized to provide conceptual models that influence the perception of and interaction with AI systems \cite{10.1145/3635636.3664250, 10.1145/3313831.3376275, 10.1145/3613904.3642697}. Social media, in particular, contains intuitive interaction concepts that facilitate social interaction. When used as a metaphor, it offers a familiar language for modeling and communicating the affordances \cite{10.1145/301153.301168} of a system that naturally fosters networked connections, polarized identity expression, and relational communication—all of which reflect our design goals. 

Specifically, social media offers unique features that allow people to easily connect with individuals beyond their usual acquaintances, creating an environment where everything becomes a potential point of connection \cite{papacharissi2010networked}. This characteristic aligns with \textbf{DG 1}. Drawing inspiration from this, we designed FRIENDS DISCOVERY, a feature that allows writers to specify the types of connections (e.g., Harry’s long-lost friend) they wish to seek for a given character. Subsequently, the feature offers character candidates with relevant ties grounded in the defined connection point and the seed character’s personal details. 

Social media often reveals personal details about people’s lives, helping to build social bonds and reinforce a sense of identity \cite{10.1145/2317956.2317994}. Furthermore, when anonymity is guaranteed, individuals are more likely to engage in self-disclosure \cite{pohan2022use}. This often results in a shared environment where honest and at times even polarizing \cite{lee2023makes} reactions to a single event are openly expressed and compared. These traits match \textbf{DG 2}. Inspired by this, we designed JOURNALS, a feature that allows writers to first define a theme (e.g., Ron has vanished) and then generate diary entries in which selected characters articulate their inner thoughts on that shared topic. 

Social media is frequently used to exhibit and strengthen relationships between users. For instance, comment threads commonly serve as a space to express interpersonal affection, remark on some mutual events, or proclaim relational status \cite{ballantine2015influence}. These characteristics map onto \textbf{DG 3}. Based on this, we designed COMMENTS, a feature in which writers can select characters to freely respond to each other’s journal entries, allowing their relationships to unfold through open-ended interactions generated by this feature. 

Lastly, across these features, outputs are presented in formats (e.g., diary entries) aligned with their respective interaction concepts. By presenting text in forms distinct from conventional narrative prose, we encourage writers to treat the generated content as raw materials to adapt before incorporating them into the narrative, if incorporation happens at all. This approach aims to render the integration of AI-generated content as a deliberate, transformative act and to ensure that writing the final manuscript remains in the writer’s own hand, responding to prior calls to carefully consider authorial agency in the design of AI-assisted writing tools \cite{carrera2025add}. 

\subsection{System Components}
\label{sec:4.2}
The social media metaphor, adopted in response to our design goals, also serves as the basis for Constella’s interface design. We now describe its interface components along with the social media-inspired interaction concepts they realize.

\begin{figure}[H]
    \centering
    \includegraphics[width=0.95\textwidth]{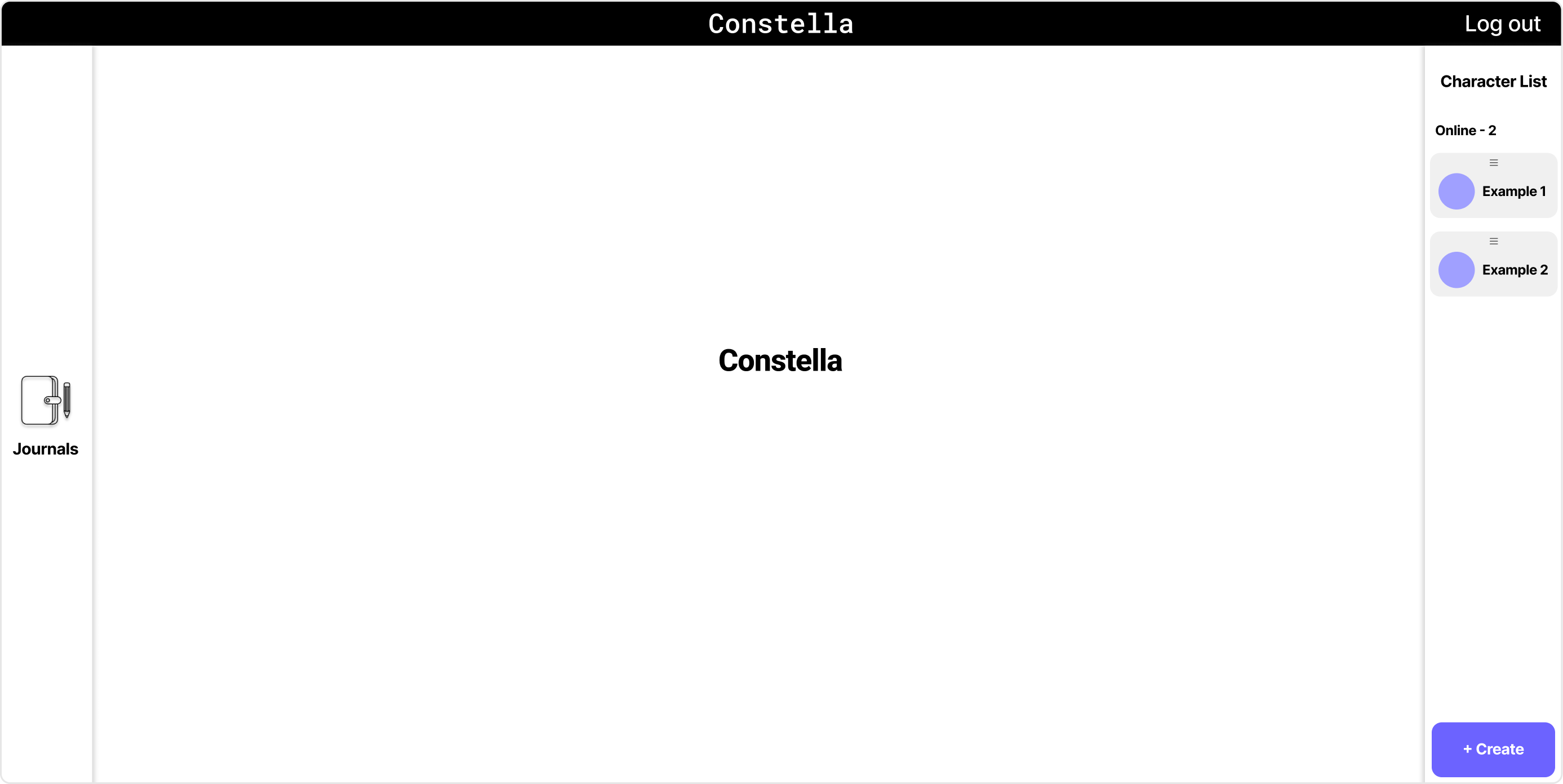}
    \caption{Constella’s default layout. The Left Sidebar opens the Journals Panel. The Right Sidebar shows a list of Character Cards, each of which opens the Profile Panel. The Create Button opens the New Character Panel.}
    \Description{Constella’s default layout includes a ‘Log out’ button, a left sidebar labeled ‘Journals,’ a large space in the middle, a right sidebar showing a ‘Character List’ with two example character cards, and a ‘Create’ button in the bottom-right corner.}
    \label{fig:fig4}
\end{figure}

The interface features a blank canvas (\autoref{fig:fig4}) where users can horizontally stack and reorder multiple panels. There are three panel types: New Character Panel (§\ref{sec:4.2.1}), Profile Panel (§\ref{sec:4.2.2}), and Journals Panel (§\ref{sec:4.2.3}).  

    \subsubsection{New Character Panel}
    \label{sec:4.2.1}
    
    \begin{figure}[H]
        \centering
        \includegraphics[width=0.95\textwidth]{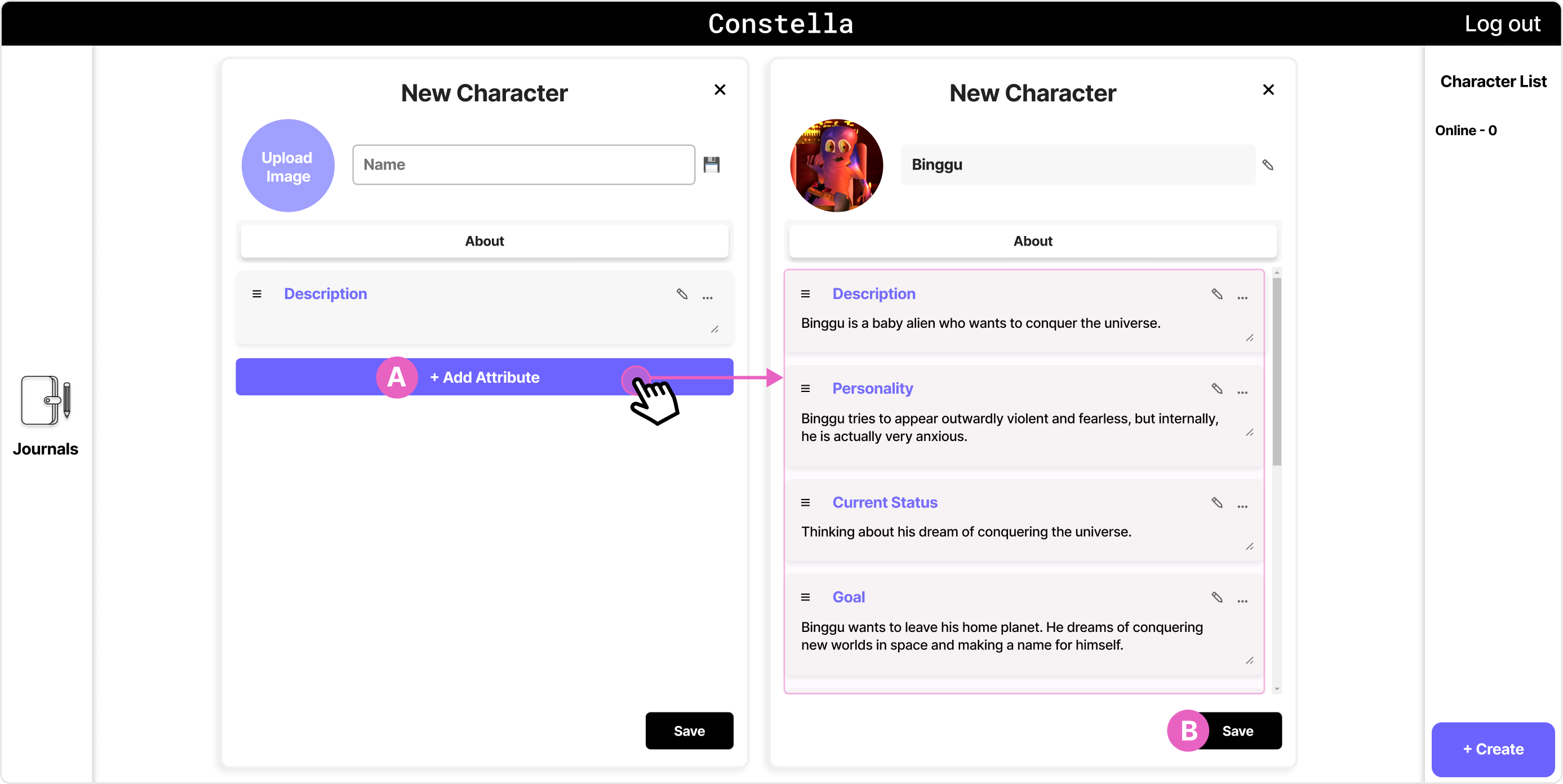}
        \caption{New Character Panel. (A) Add Attribute Button. (B) Save Button.}
        \label{fig:fig5}
        \Description{Two New Character Panels are shown side by side. The left panel shows the initial state, and the right panel shows the filled-in state with many added attributes for Binggu. The "Add Attribute" button and "Save" button are located within these panels.}
        
    \end{figure}
     In this panel, users can create new characters by setting a name, uploading an image, and adding customizable Attributes (e.g., age, gender, and backstory) (\autoref{fig:fig5}-A). Attributes can be edited, deleted, or reordered. Once characters are saved (\autoref{fig:fig5}-B), their profiles can be accessed from the Right Sidebar.
     
    \subsubsection{Profile Panel}
    \label{sec:4.2.2}
    
    \begin{figure}[H]
        \centering
        \includegraphics[width=0.95\textwidth]{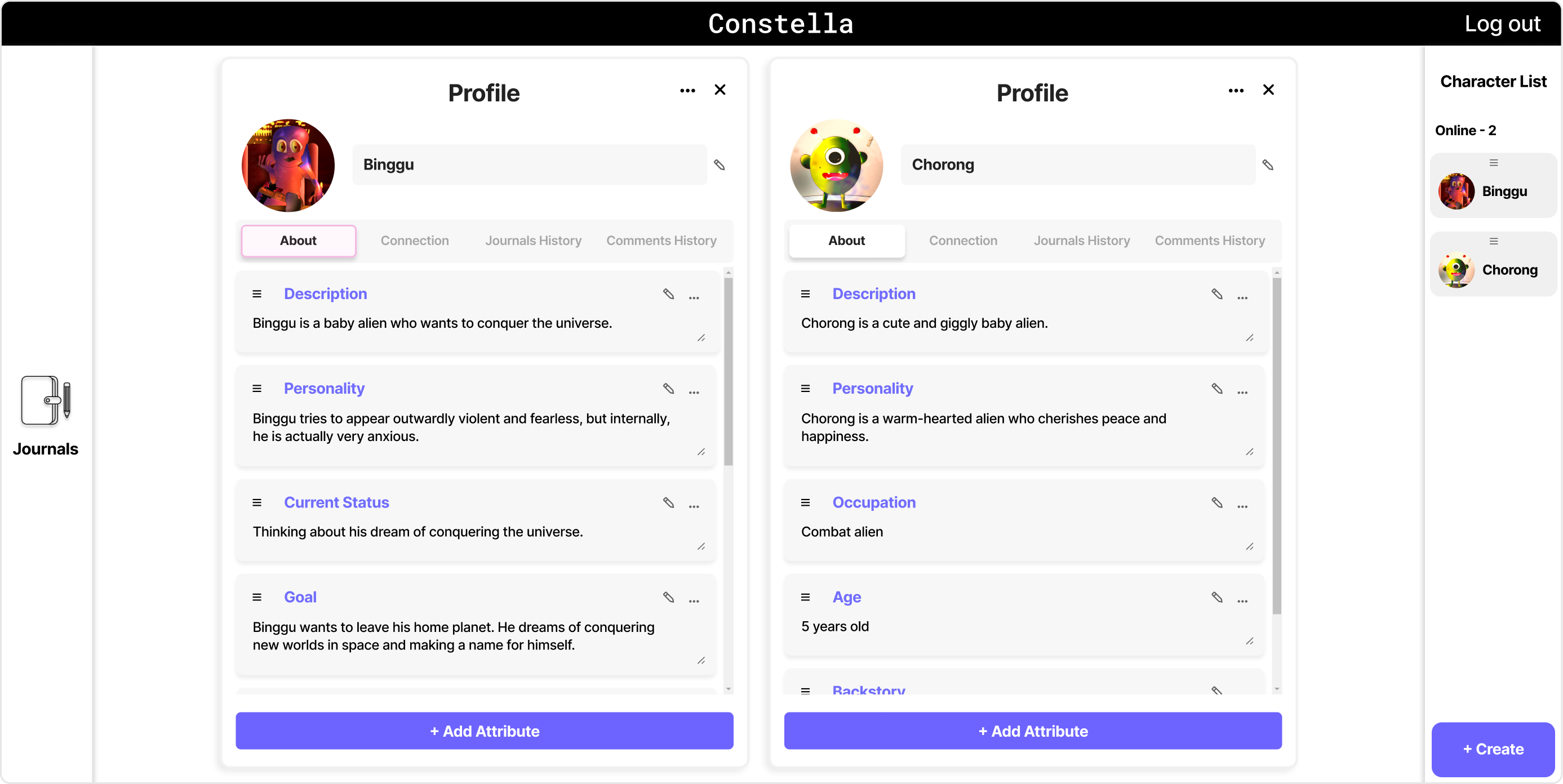}
        \caption{Profile Panel’s About Tab.}
         \Description{Two Profile Panels are shown side by side, both set to the About Tab. The left panel is for Binggu, and the right panel is for Chorong.}
        \label{fig:fig6}
    \end{figure}

    \begin{figure}[H]
        \centering
        \includegraphics[width=0.95\textwidth]{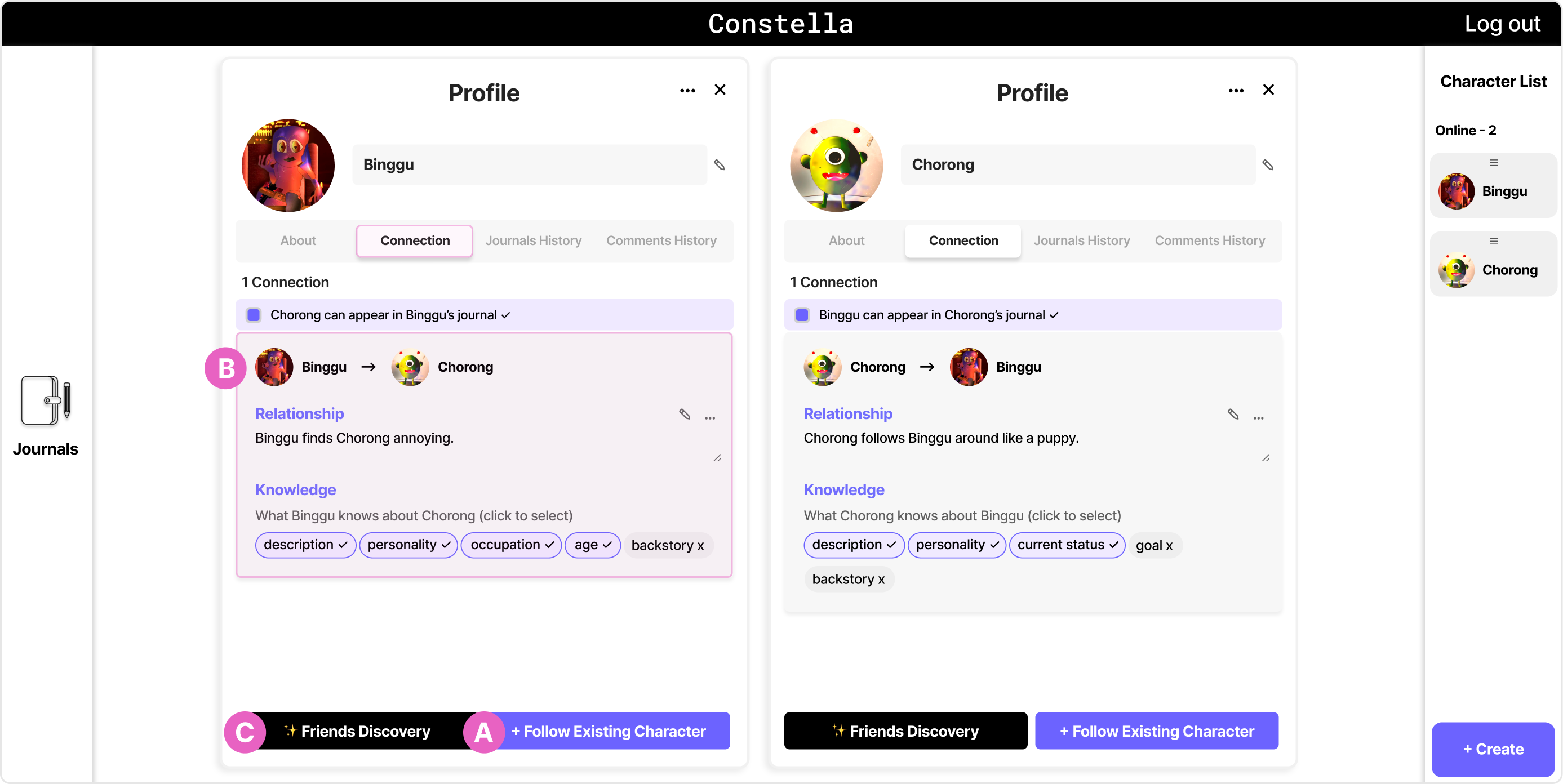}
        \caption{Profile Panel’s Connection Tab. (A) Follow Existing Character Button. (B) Relationship Attributes. (C) Friends Discovery Button.}
        \Description{Two Profile Panels are shown side by side, both set to the Connection Tab. The left panel is for Binggu, and the right panel is for Chorong. The "Follow Existing Character" button, relationship attributes, and "Friends Discovery" button are located within this tab.}
        \label{fig:fig7}
    \end{figure}
    The Profile Panel has four tabs: 1) About Tab, 2) Connection Tab, 3) Journals History Tab, and 4) Comments History Tab. 
    
    In the About Tab, users can continue to edit Attributes (\autoref{fig:fig6}). In the Connection Tab, users can make the current character (Binggu) follow other characters (Chorong) (\autoref{fig:fig7}-A) to create Relationship Attributes (\autoref{fig:fig7}-B). Within the Relationship Attributes, users can specify relational details, such as how the current character perceives the other character or events that occurred between them. Users can also manage the current character’s Knowledge of others. For instance, the current character might know about the other’s occupation but not their backstory. Relationship Attributes can be modified at any time. 

    \begin{figure}[H]
        \centering
        \includegraphics[width=0.95\textwidth]{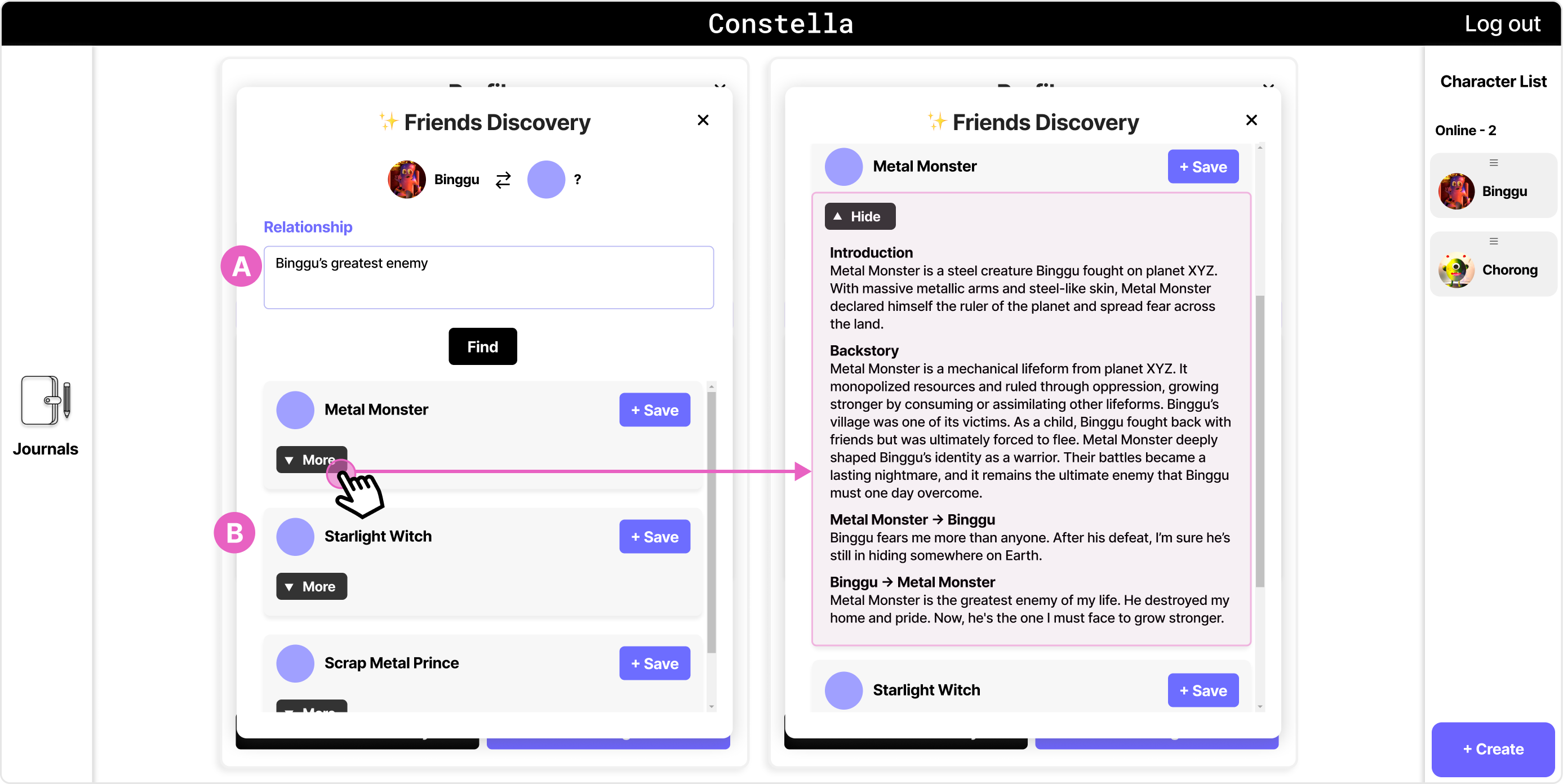}
        \caption{FRIENDS DISCOVERY feature. (A) Prompt input field. (B) Generated Mini Profiles.}
        \Description{Two Profile Panels for Binggu are shown side by side, both with the Friends Discovery pop-up window open. The left panel’s pop-up shows a relationship prompt input field and three generated mini profiles, each with a "More" option and a "Save" button. The right panel’s pop-up shows the expanded details for a generated character, Metal Monster.}     
        \label{fig:fig8}
    \end{figure}

    Additionally, by clicking the Friends Discovery Button (\autoref{fig:fig7}-C) and entering a relational prompt (\autoref{fig:fig8}-A), users can generate Mini Profiles for three new characters related to the current character (\autoref{fig:fig8}-B).

    \begin{figure}[H]
        \centering
        \includegraphics[width=0.95\textwidth]{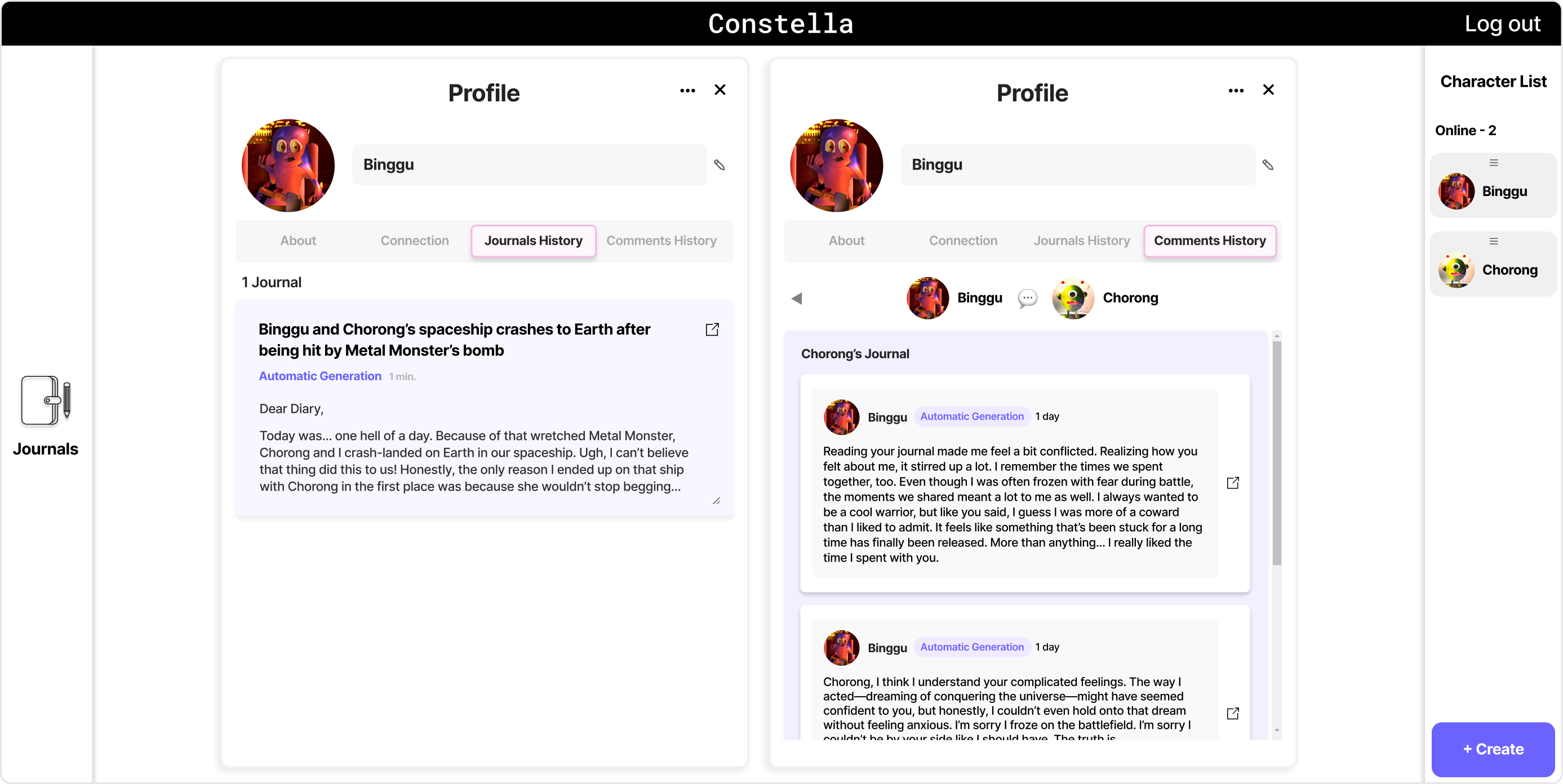}
        \caption{Profile Panel’s Journals History Tab (Left) and Comments History Tab (Right).}
        \Description{Two Profile Panels for Binggu are shown side by side. The left panel is set to the Journals History Tab, and the right panel is set to the Comments History Tab.}
        \label{fig:fig9}
    \end{figure}

    The Journals History Tab displays all journal entries created by the current character, and the Comments History Tab shows all comment exchanges involving that character (\autoref{fig:fig9}). All characters saved in the system can also be deleted by the user at any time.

\clearpage 

    \subsubsection{Journals Panel}
    \label{sec:4.2.3}
    
    \begin{figure}
        \centering
        \includegraphics[width=0.95\textwidth]{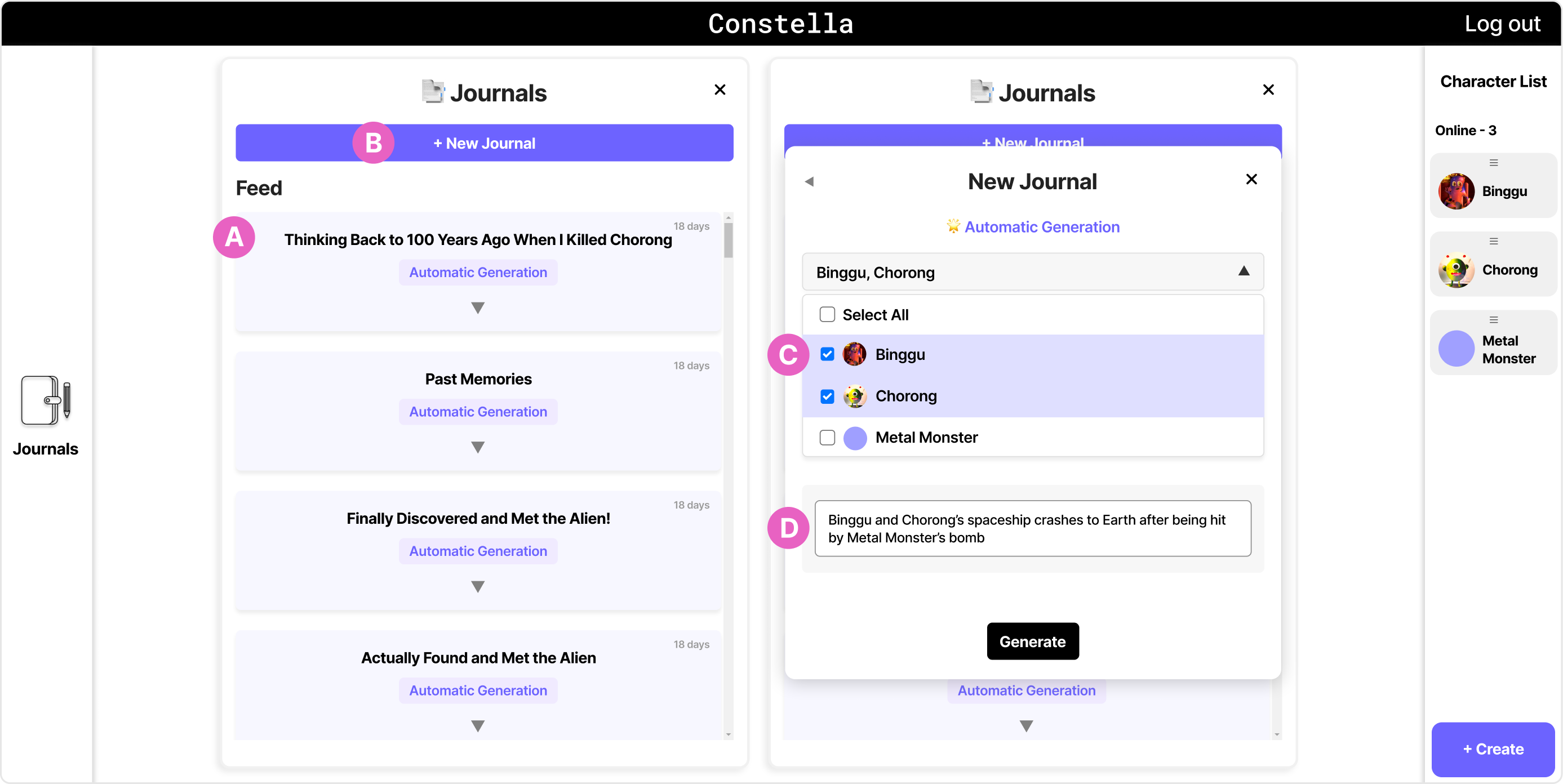}
        \caption{Journals Panel. (A) Collection of previously generated journal entries. (B) New Journal Button. (C) Character selection field. (D) Prompt input field.}
        \Description{Two Journals Panels are shown side by side. The left panel shows a feed of previously generated journal entries and a "New Journal" button at the top. The right panel shows the pop-up window that appears when the "New Journal" button is clicked, with a character selection field, a prompt input field, and a "Generate" button.}
        \label{fig:fig10}
    \end{figure}

    \begin{figure}
        \centering
        \includegraphics[width=0.95\textwidth]{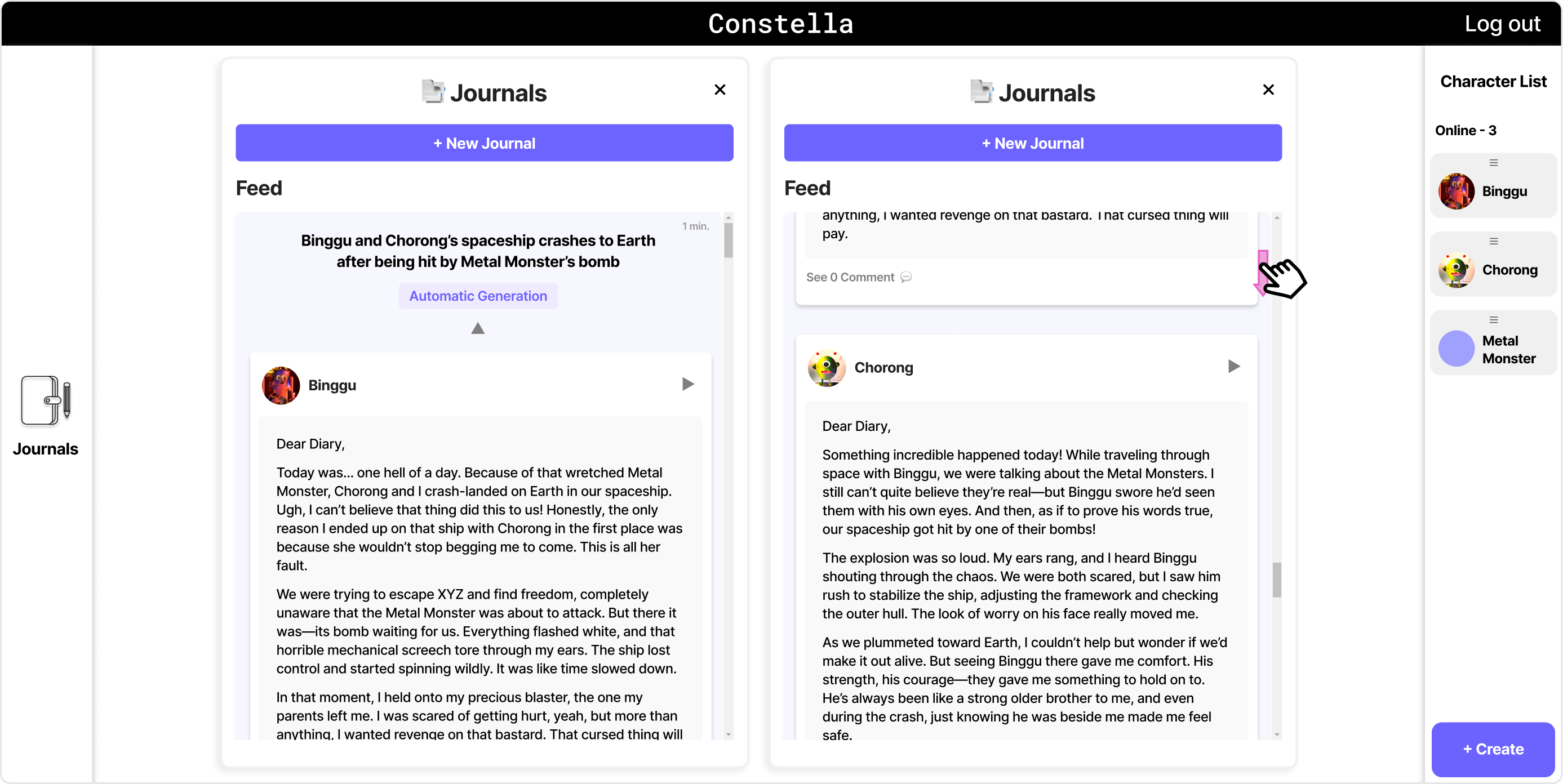}
        \caption{Automatically generated journal entries by two characters based on the same theme. The second journal entry can be viewed by scrolling down from the first generated journal entry.}
         \Description{Two Journals Panels are shown side by side, displaying generated journal entries. The left panel shows Binggu’s journal entry, and the right panel shows Chorong’s journal entry.}
        \label{fig:fig11}
    \end{figure}
    
In the Journals Panel, users can view existing journal entries (\autoref{fig:fig10}-A) or create new ones (\autoref{fig:fig10}-B) through two modes. In Manual Writing Mode, users select characters and manually write both the theme and content of the journal entries. In Automatic Generation Mode, users can select any number of characters (\autoref{fig:fig10}-C) and write their own theme (\autoref{fig:fig10}-D) to produce generated entries for each character (\autoref{fig:fig11}). While the theme may take the form of a situation, an emotional state, or even a philosophical inquiry, the system ultimately treats it as a topic for the characters to write about. All journal entries are editable and deletable. 

\clearpage 
    \begin{figure}
        \centering
        \includegraphics[width=0.95\textwidth]{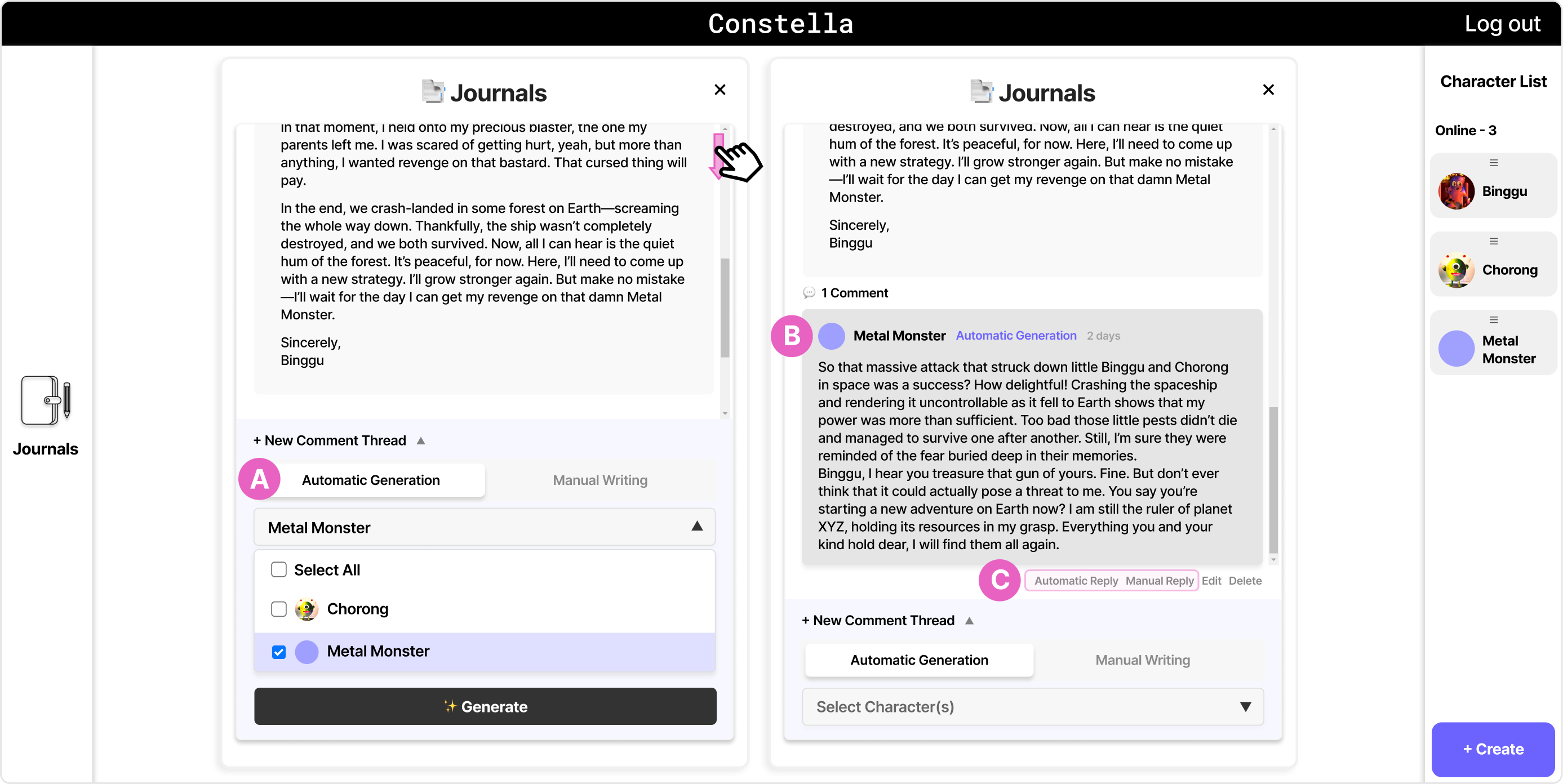}
        \caption{COMMENTS view. (A) Create a new comment thread. (B) A generated comment thread. (C) Create a reply to a comment.}
          \Description{Two Journals Panels are shown side by side, both set to the Comments view. The left panel shows the interface for creating a new comment thread, including a mode-selection tab, a character selection field, and a "Generate" button. The right panel shows a generated comment thread by Metal Monster, with buttons underneath for automatic reply, manual reply, edit, and delete.}
          
        \label{fig:fig12}
    \end{figure}
The Comments view below each journal entry can be used to create new comment threads (\autoref{fig:fig12}-A) in two modes. In Manual Writing Mode, users select characters and manually write comments from their perspective. In Automatic Generation Mode, users select characters to generate comments in response to the journal entry. Users can also add new comments to existing threads (\autoref{fig:fig12}-B) using either Automatic Reply or Manual Reply Modes (\autoref{fig:fig12}-C).

Any character other than the journal’s author can be selected to initiate a new comment thread. The journal author and the selected commenter can then take alternating turns to reply to each other’s comments within that thread. This structure allows for a broader range of reactions to the journal entry without compromising the depth that comes from focused one-on-one interactions. All comments are editable and deletable.

\subsection{An Example Use Scenario}
\label{sec:4.3}
To demonstrate our system, we present how Jina, an imaginary animation writer we created for this example, uses Constella for her project, "The Conqueror." We note that this is simply one possible use case, rather than a prescriptive or optimal workflow.

\subsubsection{Character Profiles}
\label{sec:4.3.1}
Jina has roughly conceptualized two of her main characters, "Binggu" and "Chorong." She opens two New Character Panels (\autoref{fig:fig5}) in Constella and adds several Attributes (e.g., age, goal, backstory, personality) for each character. For Binggu’s personality, she writes, "Binggu tries to appear outwardly violent and fearless, but internally, he is actually very anxious." For Chorong, she writes, "Chorong is a warm-hearted alien who cherishes peace and happiness." She saves both characters. 
Next, Jina opens their Profile Panels (\autoref{fig:fig6}) and creates their Relationship Attributes by making them follow each other (\autoref{fig:fig7}). Within Binggu’s profile, she writes, "Binggu finds Chorong annoying," and decides that Binggu knows Chorong’s character description, personality, occupation, and age. Within Chorong’s profile, she states, "Chorong follows Binggu around like a puppy" and specifies that Chorong knows Binggu’s character description, personality, and current status.

\subsubsection{FRIENDS DISCOVERY}
\label{sec:4.3.2}
Jina now thinks it would be intriguing to have these two characters crash-land on Earth from outer space. To create a character who causes this crash, she uses FRIENDS DISCOVERY. She enters "Binggu’s greatest enemy" (\autoref{fig:fig8}-A) as the input and receives three suggestions: "Metal Monster," "Starlight Witch," and "Scrap Metal Prince" (\autoref{fig:fig8}-B). After browsing through the Mini Profiles, Jina is drawn to Metal Monster. As he was the one who wrecked Binggu’s home in the past, Metal Monster would be the perfect villain to stir up Binggu’s emotional journey. Without hesitation, Jina locks him in as the story’s antagonist by saving the character.

\subsubsection{JOURNALS}
\label{sec:4.3.3}
Jina is now curious about how Binggu and Chorong would react during their crash landing on Earth. Using JOURNALS, she selects both characters (\autoref{fig:fig10}-C) and generates their journal entries by entering the theme "Binggu and Chorong’s spaceship crashes to Earth after being hit by Metal Monster’s bomb" (\autoref{fig:fig10}-D). Through the generated entries (\autoref{fig:fig11}), Jina discovers that Binggu experiences overwhelming stress and frustration in the moment of the blast. Meanwhile, Chorong views the situation with a hint of optimism, tinged even with admiration for the surreal experience. Inspired by this contrast, Jina decides to further amplify their differences by making Chorong even more naive and filled with wonder. Chorong would then perhaps even marvel at the sparkly colors of Earth’s atmosphere during the crash and express excitement about the unexpected adventure that lies ahead. Jina updates Chorong’s Attributes to reflect these changes. 

\subsubsection{COMMENTS}
\label{sec:4.3.4}
Jina is now eager to explore how Metal Monster would respond to Binggu’s journal entry. Using COMMENTS, Jina generates Metal Monster’s comment (\autoref{fig:fig12}-A) and finds that Metal Monster is still harboring resentment toward Binggu even after his bomb landed successfully (\autoref{fig:fig12}-B). Seeing this, Jina starts to wonder what unresolved emotions are fueling his hostility toward Binggu. Motivated by this question, she updates Metal Monster’s Relationship Attributes, adding details about the simmering jealousy that has quietly shaped his attitude toward Binggu all along.
\subsubsection{Future Use}

\label{sec:4.3.5}
In the future, Jina can discover new friends Binggu meets on Earth using FRIENDS DISCOVERY, investigate how Metal Monster and Chorong would each react when tasting candy for the first time through JOURNALS, and witness Chorong’s teasing interrogation of Metal Monster’s journal entry through COMMENTS. Example generation results for each feature, based on this future usage, are also included in Appendix \ref{sec:Appendix-prompts}.

\subsection{Implementation and Prompt Design}
\label{sec:4.4}
    \subsubsection{Implementation}
    \label{sec:4.4.1}
    Constella was developed as a web application with a React.js front-end and a Node.js back-end. User-uploaded images were stored in Google Cloud Storage, while all other application data were managed in MongoDB. 

    As the participants in our user study were South Korean, we aimed to use an LLM with strong proficiency in both Korean language and cultural understanding. At the time of our research, however, there were no widely published evaluations of state-of-the-art LLMs, such as OpenAI’s GPT-4o \cite{hurst2024gpt} and the Claude 3 family \cite{anthropic2024claude3}, with regard to Korean-specific performance. To address this, we referred to a publicly available community report \cite{corca2024gpt4o_click}, which evaluated GPT-4 and GPT-4o on CLIcK \cite{kim-etal-2024-click}, a dataset specifically designed to assess cultural and linguistic intelligence in Korean. Their results showed that GPT-4o outperformed GPT-4 and other models tested in the original CLIcK paper. Based on this, we selected GPT-4o and accessed it via the OpenAI API.
    
    \subsubsection{Prompt Design}
    \label{sec:4.4.2}
Our prompt design was informed by OpenAI’s prompt engineering guide \cite{openai2025textgen} and prior research \cite{10.1145/3613904.3642545, 10.1145/3613904.3642406, 10.1145/3613904.3642472} that disclosed prompts for persona-based roleplaying. This section outlines the key elements of the prompts used for each AI feature in our system, focusing on their static instructions, provided context, and expected outputs (see \autoref{tab:prompt-design} for a summary). We provide the full set of prompts in Appendix \ref{sec:Appendix-prompts}.

\begin{table}[ht]
\renewcommand{\arraystretch}{1.2}
\caption{Components of each feature’s prompt design.}
\label{tab:prompt-design}
\resizebox{\textwidth}{!}{
\begin{tabular}{p{1.8cm} p{4.5cm} p{2.8cm} p{3.25cm} p{3.4cm} p{2.05cm}}
\toprule
\textbf{Feature} & \textbf{Static Instruction} & \multicolumn{3}{c}{\textbf{Provided Context}} & \textbf{Output} \\
\cmidrule(lr){3-5}
 &  & \textbf{User Input} & \textbf{Character Profile} & \textbf{Generation History} &  \\
\midrule
\raggedright FRIENDS\\[-0.5ex]\raggedright DISCOVERY & \raggedright To roleplay as a creative writer. Rules for character generation. & \raggedright Relationship phrase \newline & Attributes & - & Mini profiles \\
\midrule
\raggedright JOURNALS & \raggedright To roleplay as a character. \newline Rules for journaling. & \raggedright Journal theme \newline & Attributes, \newline Relationship Attributes & - & Journal entries \\
\midrule
\raggedright COMMENTS & \raggedright To roleplay as a character. \newline Rules for commenting. & - & Attributes, \newline Relationship Attributes & Journal entry, \newline (if any) prior comments & Comments \\
\bottomrule
\end{tabular}}
\end{table}

\paragraph*{\textbf{FRIENDS DISCOVERY}} The static instruction guides the model to roleplay as a storywriter and generate three distinct fictional characters based on a specified relationship phrase and the given character’s Attributes. Each generated character is composed of a name, introduction, backstory, and bidirectional relationship descriptions. We deliberately chose to provide a minimal set of Attributes following \textbf{DG 1}. The outputs are formatted as JSON objects to ensure proper display as Mini Profiles within the user interface. 

\paragraph*{\textbf{JOURNALS}} The static instruction directs the model to roleplay as the given character, using both the provided character’s Attributes and Relationship Attributes. The model then generates a journal entry based on a specified theme. In alignment with \textbf{DG 2}, each entry begins with the phrase "Dear Diary," thereby establishing the entry as an intimate monologue that reveals the character’s inner self to an imagined confidant. When multiple characters are selected, this process is executed in parallel, and the resulting entries are presented together for simultaneous viewing. 

\paragraph*{\textbf{COMMENTS}} The static instruction places the model in the character’s role by making it refer to the character’s Attributes and Relationship Attributes. Taking into account the open-endedness emphasized in \textbf{DG 3}, the model is given freedom to focus on a specific aspect of the journal entry when generating a response. In a threaded reply, the model is further guided by the preceding comment history to maintain continuity while focusing on the most recent comment in the thread. 

In addition, our prompt design adopts a stateless generation approach; characters do not retain memory of previously generated journal entries beyond the current one, nor of comments outside the current thread. This design choice was motivated by the assumption that character Attributes would be iteratively revised during writers’ creative processes. Persisting prior outputs could thus introduce inconsistencies with the updated character definitions and potentially burden users to micromanage previously generated content. Thus, we believed that a deliberate forgetting strategy \cite{10.5555/AAI29327510} would help writers iterate more freely in favor of emergent ideas and reduce creative pressure by eliminating a sense of finality.

%
\section{User Study}
\label{sec:5}
We conducted a 7–8 day system deployment study to evaluate Constella. Our aim was to understand how Constella integrates into storywriters’ existing practices and how it supports their interconnected character creation process. 

CSTs are often evaluated through a mix of observing user actions during the creative process and subjective experience reports \cite{10.1145/3635636.3656204} (e.g., Creativity Support Index (CSI) \cite{10.1145/2617588}). Given that Constella was designed to assist storywriters’ creative processes within the context of full-length story creation, we chose to evaluate our tool by exploring its use through a system deployment study. This approach provides participants with the temporal and spatial flexibility to use the tool over an extended period within their natural environments and established processes \cite{10.1145/3357236.3395474}. To closely track their experiences throughout this period, we collected system logs that stored all our participants’ actions (e.g., generations, edits) performed within Constella. We also employed a diary study, along with follow-up interviews, to revisit participants’ experiences in greater depth. Lastly, we selected the CSI to gather quantifiable subjective experience reports. The CSI is a widely used and psychometrically validated survey instrument that measures how well a system supports creative activities. We specifically used the survey to identify the strengths and weaknesses of Constella in its role as a CST. The study process was approved by the university’s Institutional Review Board (IRB No. KH2024-140).

\subsection{Participants}
\label{sec:5.1}

\begin{table}[ht]
    \renewcommand{\arraystretch}{1.4}
    \caption{User study participants’ self-reported demographics}
    \label{table:userstudy-demo}
    \resizebox{\textwidth}{!}{%
    \begin{tabular}{@{}lll m{3cm} m{4cm} m{3.1cm}
                m{1.6cm} m{1.6cm} m{1.8cm}@{}}
    \toprule 
    \textbf{ID} & \textbf{Gender} & \textbf{Age} & \textbf{Story Types} & \textbf{Genres} & \raggedright \textbf{Years of Creative}\\[-0.5ex]\raggedright \textbf{Writing Experience} & \raggedright \textbf{Related}\\[-0.5ex]\raggedright \textbf{Degree} & \raggedright \textbf{Publishing}\\[-0.5ex]\raggedright \textbf{Experience} &\makecell[l]{\textbf{LLM}\\[-0.5ex]\textbf{Experience}} \\
    \midrule
    P1  & Female & 31 & TV Show, Novel & Mystery, Romance, Comedy & 15 & Bachelor’s & Yes & Average \\
    P2  & Female & 22 & Novel & Literary Fiction & 1 & None & No & Average \\
    P3  & Male & 30 & Film, TV Show & Fantasy, Science Fiction, Drama & 10 & Bachelor’s & Yes & Average \\
    P4  & Female & 34 & Novel & Fantasy, Crime, Slice-of-life, Thriller & 1 & None & No & Low \\
    P5  & Female & 33 & TV Show, Film, Novel & Romance, Literary Fiction & 15 & Master’s & No & Very High \\
    P6  & Female & 39 & Novel & Modern Romance, Eastern Fantasy & 14 & Bachelor’s & Yes & High \\
    P7  & Male & 48 & TV Show, Film & Historical Fiction, Spy Action, Melodrama & 20 & Bachelor’s & Yes & Very High \\
    P8  & Male & 32 & Comics & Fantasy, Action & 10 & None & Yes & None \\
    P9  & Female & 27 & Film, TV Show & Drama, Thriller & 5 & Bachelor’s & No & Low \\
    P10 & Male & 51 & Novel, Comics, Animation & Modern Fantasy, Kid Adventure & 20 & None & Yes & Average \\
    P11 & Female & 29 & Play, TV Show & Comedy, Fantasy, Slice-of-life, Revenge Drama, Noir, Drama & 8 & Bachelor’s & Yes & Average \\
    \bottomrule
    \end{tabular}}
\end{table}

We recruited participants with prior experience in creating full-length fictional stories and aimed to capture insights across diverse story types, genres, and creative writing experiences. We recruited participants with and without formal creative writing education, as well as those with and without publication experience. Additionally, writers who had participated in the formative study were excluded from the recruitment pool. This was to explore the broad applicability of our tool with writers whose practices may differ from those of the formative study participants. Recruitment channels and the sampling method were identical to our formative study.

As 2 out of the 13 participants withdrew from the study due to personal reasons, we report details (\autoref{table:userstudy-demo}) for our final group of 11 storywriters (F=7, M=4). Story types ranged from TV shows (N=6), novels (N=6), films (N=4), comics (N=2), plays (N=1), and animation (N=1). The total number of writing genres they had experience with reached 20. Years of creative writing experience ranged from 1 to 20 years, with an average of 10.82 years. While not part of the inclusion criteria, participants also reported their levels of experience with LLMs, as their proficiency in prompting could potentially influence how they interact with the tool. All participants resided in South Korea. 

\subsection{Procedure}
\label{sec:5.2}
    \subsubsection{Tasks}
    \label{sec:5.2.1}
    Creating a full-length story is an extensive process that can take several months or even years. Therefore, we split the process into a set of tasks focused enough to explore system usage while aiming to align with participants’ usual writing practices. 

    \paragraph*{\textbf{(1) Characters}} We asked participants to create at least five characters (profiles) within Constella. Although the number of characters in fictional stories varies widely, Truby \cite{truby2008anatomy} proposed that characters can be classified based on their roles: the hero, opponent, ally, fake-ally opponent, and subplot character. While not all stories require characters for each of these five roles, some may feature multiple characters fulfilling the same role. Thus, we did not prescribe specific character roles but requested a minimum number of characters. 

    \paragraph*{\textbf{(2) Character Backstory}} We asked participants to write a central character’s backstory. This is a foundational activity according to storywriting literature (§\ref{sec:2.1}) and one that was also utilized by our formative study participants (§\ref{sec:3.5.2}). Given that relationships often form the basis of backstories \cite{barker2019build}, we asked participants to include how their central character influenced or was influenced by other characters in their backstory. For this task, we asked writers to use FRIENDS DISCOVERY, as this feature was designed to support writers in discovering related characters. We included this task to ensure FRIENDS DISCOVERY could be utilized non-intrusively, eliminating the need for participants to incorporate entirely new characters in their narratives, as they may have already set clear goals for their character cast. In addition, we only asked participants to write a backstory for one character, as this task was sufficient for trying FRIENDS DISCOVERY. 

   \paragraph*{\textbf{(3) Scenes}} Participants were asked to write three types of scenes: Scene Type A, B, and C.

    \textit{\textbf{Scene Type A}.} A scene where several characters show contrasting or aligned perspectives. The former can be interpreted as conflict and the latter as unity, both essential elements in stories \cite{egri1972art}. Therefore, we believed that this scene could naturally be integrated into their narrative. We encouraged participants to use JOURNALS when writing this scene, as the feature is intended to help writers delve into and compare their characters’ thoughts, making it relevant to this task.

     \textit{\textbf{Scene Type B}.} A scene that shows the relationship between two characters. As character relationships are also central components in stories, we expected these requirements to cause minimal deviation from their narrative goals. We asked writers to use COMMENTS when writing this scene, as this feature was designed to help writers elaborate on character relationships. We also note that while participants could include any number of multiple characters per scene, we asked them to focus on two-character relationships in Scene Type B, as COMMENTS mainly supports dyadic interactions (§\ref{sec:4.2.3}).

     \textit{\textbf{Scene Type C}.} A scene that involved several characters. For this task, participants were informed to use at least one of the system’s AI features. We provided minimal requirements for this scene to explore additional system usage when writers were given more freedom to use Constella according to their own needs. 

    \paragraph*{\textbf{(4) Story Outline}} We asked participants to complete a story outline for a full-length story featuring at least five characters (profiles). Before completion, we asked writers to use at least one of Constella’s AI features. This task was included because this study aimed to explore system usage within the context of writing a full-length story. The act of writing a story outline provided this overarching context, while its compressed nature, such as lacking details compared to the entire full-length story, was complemented by the aforementioned tasks. 

    \subsubsection{Supplementary Task Guidelines and Ethical Protocols}
    \label{sec:5.2.2}
    We provided supplementary guidelines to support task flexibility, accommodate participants’ preferred writing practices, and ensure ethical handling of intellectual property.

    \textit{\textbf{System Outcomes}.} Participants were asked to try the system’s features for the tasks (§\ref{sec:5.2.1}), but the decision to actually use the system-generated outcomes and how to incorporate them was entirely at their discretion.

    \textit{\textbf{Order}.} The aforementioned tasks are not sequential steps but a culmination of final results. For example, participants could begin outlining their story while creating character profiles, write Scene Type C, and then revise their outline. Consequently, there was no strict order in which they had to use the system’s features.

   \textit{\textbf{Length and Format}.} Participants could freely select the story type and genre for their narrative, which ultimately influenced the length and format of the tasks. For instance, some writers preferred a screenplay format, often writing more pages due to dialogue and spacing conventions. Others favored detailed prose, which tended to be more textually dense despite occupying fewer pages. 

    \textit{\textbf{Word Processor}.} Participants selected their preferred word processor (e.g., Microsoft Word) to complete the tasks.

    \textit{\textbf{Intellectual Property}.} Participants were informed, both during recruitment and before the start of the study, that their work (backstory, scenes, outline) would not be disclosed in its entirety to external parties. Instead, it would be used solely to provide context during our data analysis process. However, we clarified that all content created within Constella, whether generated or manually written, could be fully disclosed for academic purposes. All participants agreed to the above terms. To formalize this process, we provided a non-exclusive intellectual property license agreement detailing these terms, which each participant chose to sign.

    \subsubsection{Study Procedure}
    \label{sec:5.2.3}

    \begin{figure}[H]
        \centering
        \includegraphics[width=\textwidth]{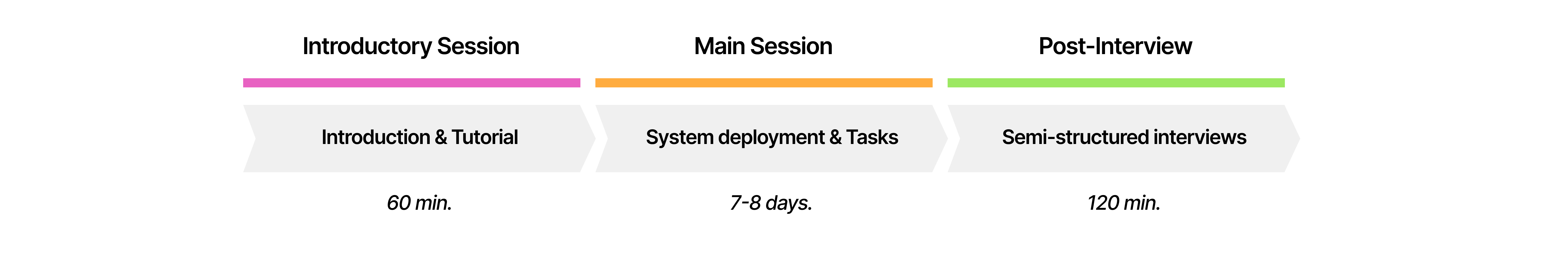}
        \caption{Procedure of the user study.}
        \Description{Fully described in the text.}
        \label{fig:userstudy-process}
        
    \end{figure}
    The user study consisted of three parts: (1) introductory session, (2) main session, and (3) post-interview (\autoref{fig:userstudy-process}). We describe each part in detail below.

    \textbf{(1) Introductory Session (60 minutes)}: We conducted individual introductory sessions with each participant via Zoom. After obtaining informed consent, we introduced the goal of our study and provided a live system tutorial. To offer an overview of the system’s features, the live tutorial included the first author demonstrating a use case largely following §\ref{sec:4.3} while explaining the intended purpose and motivation for each feature. Participants then engaged in a hands-on session to explore the tool, using each of Constella’s main features by creating their own examples. Specifically, we asked participants to create two rough character profiles, which could be based on people they already know or characters they had previously created to make the process less burdensome. We then asked them to use the FRIENDS DISCOVERY, JOURNALS, and COMMENTS features a couple of times. Participants were also encouraged to ask questions during both the live tutorial and the hands-on session. After this process, we provided guidelines for the diary study and explained the tasks for the main session. 

    \textbf{(2) Main Session (7–8 days)}: Over a 7–8 day period, participants were asked to complete the assigned tasks and record their thoughts in an online diary each day. For the diary, we utilized Notion \cite{notion2025} to enable quick and easy documentation. The diary included questions (\autoref{table:userstudy-diary} in Appendix \ref{sec:Appendix-questions}) to guide participants in documenting both their storywriting process and system usage experience. While our initial timeframe for the study was set at 7 days, a duration determined through a pilot study, six participants requested an additional day to complete their tasks. As a result, five participants took part in the main session for 7 days, and six participants for 8 days. 

    \textbf{(3) Post-Interview (120 minutes)}: Finally, we conducted individual semi-structured interviews with each participant to gain an in-depth understanding of their experiences. Each interview was conducted via Zoom. We began by asking about their overall experience with the system, followed by specific questions about each feature (\autoref{table:userstudy-interview} in Appendix \ref{sec:Appendix-questions}). We then inquired about their Notion diary entries to understand the context and meaning behind their documented responses. Throughout the interview, participants were encouraged to reference their Notion diaries, the system, and their completed tasks to recall and explain their experiences in detail. At the end of this session, participants completed the CSI. All participants received compensation of 350,000 KRW (approx. 251 USD) for their participation. In addition to the tutorial and post-interview session, we expected approximately 3+ hours of engagement on each day of the main session, including reflecting on their work, using the system, writing and revising their story, and completing their daily diary. The compensation was set to reflect this expected level of participant involvement, which we anticipated would also affect their regular work activities.

\subsection{Analysis}
\label{sec:5.3}
    \subsubsection{Quantitative Analysis}
    In this subsection, we describe our quantitative analysis of system logs, supplemented by participants’ Notion diary entries and their final work. The term \textit{final work} refers to the backstory, scenes, and story outline that participants submitted at the end of the study.
    
    To first gain a high-level understanding of which AI features participants used while working on specific tasks, we combined system logs with participants’ Notion diaries. We used their Notion diaries because they provided day-by-day records of which tasks participants were working on. This resulted in \autoref{fig:timeline}. We then collected only the system logs about the profiles for the characters that appeared in their final work and the journal entries and comments that remained in participants’ Constella accounts for those characters, on the assumption that non-deletion in the system may indicate potential relevance to their final work. We give an overview of these data in §\ref{sec:6.1.1} and use them as the basis for all subsequent analyses, referring to them as \textit{system data} in this subsection.
    
    Before probing deeper into the relationship between the system data and participants’ final work, we analyzed text-level edits performed by writers within the system. This is because we viewed editing as an indicator of appropriation, a step we expected would typically occur before incorporating, if at all, the generated content into their final work. For each piece of content, we computed three complementary metrics: 1) edit counts, 2) word retention rate, and 3) word insertion rate. Edit counts reflect the volume of revision operations conducted by the user within the system. Word retention rate is the proportion of word tokens in an item’s first saved version that were preserved in the last version saved in Constella. Conversely, word insertion rate is the proportion of word tokens in the item’s last version that were not present in the first saved version. In addition, for profiles, we only counted edits to attribute and relationship attribute descriptions (see §\ref{sec:4.2} for their definitions) to exclude trivial organizational or labeling adjustments.

    Next, we examined potential links between the system data and participants’ final work. Specifically, for character profiles, we first calculated the number of times each character was mentioned in the final work, using these counts as a proxy for narrative prominence. A Spearman’s correlation\footnote{Spearman’s correlation \cite{spearman1961proof} measures how strongly two variables are related in terms of their rank order rather than their exact numerical values. It ranges from $-1$ (perfect negative association) to $1$ (perfect positive association), with $0$ indicating no monotonic relationship.} was then conducted to evaluate the relationship between the number of times each character was mentioned in the final work and the number of edits for each character profile. Through this, we examined whether characters with higher edit counts tend to be mentioned more frequently (more narratively prominent), and as our primary interest was in monotonic associations, we chose Spearman’s correlation as opposed to Pearson’s correlation\footnote{Pearson’s correlation \cite{pearson1896vii} measures the strength of a linear relationship between two numerical variables.}. For JOURNALS and COMMENTS, we proceeded to examine content reuse by identifying generated sentences that either appeared verbatim in the final work (exact substring match) or shared a Jaccard similarity coefficient\footnote{Jaccard similarity coefficient \cite{jaccard1901etude} measures how much two sets overlap. For two sets $A$ and $B$, it is defined as $J(A,B) = \frac{|A \cap B|}{|A \cup B|}$, which ranges from 0 to 1. In our case, $A$ and $B$ are the sets of words in two sentences, so a higher value means the sentences share more of the same words.} of $\ge 0.3$ with any sentence in the final work (paraphrasing with significant vocabulary retention). We selected 0.3 as the threshold because prior work found it effective for separating related versus unrelated sentences while minimizing irrelevant matches \cite{tien2018detecting}. To exclude trivial utterances, we filtered out sentences containing fewer than 3 tokens. This choice adapts a heuristic from Husain et al. \cite{husain2019codesearchnet}, who discard text snippets shorter than three tokens on the grounds that such textual content is unlikely to be informative.
    
    \subsubsection{Qualitative Analysis}
        In this subsection, we present the qualitative analysis conducted on participants’ interview data and diary entries, supplemented by their final work and system logs. 
        
        To prepare our data for analysis, we transcribed over 1,320 minutes of audio recordings from interviews with 11 participants. The interview transcriptions and diary entries were then analyzed through reflective thematic analysis \cite{braun2019reflecting}. 
        
        Our analysis process was conducted through the following procedures. First, the first and second authors repeatedly read through the entire dataset of the diary entries and interview transcriptions. Then, these two authors developed initial codes by categorizing the data based on our tool’s features. After this process, the first and second authors reviewed the initial codes while cross-referencing system logs and completed tasks to identify themes. These themes were then iteratively refined through discussions among all the authors. 
        
        As a result of this six-month process, we constructed themes related to use cases for interconnected character creation, the tool’s impact on writers and their writing processes, and its potential future usages, improvements, and limitations. We also note that we found no significant thematic variation based on participants’ duration of creative writing experience or LLM experience and thus did not pursue further analysis along those dimensions.

\section{Results}
\label{sec:6}
In this section, we report our findings by first providing an overview of results drawn from system logs and CSI results (§\ref{sec:6.1}). Next, we present several use cases that illustrate how Constella’s AI features support interconnected character creation (§\ref{sec:6.2}). We then report the tool’s impact on writers and their writing processes, with particular attention to its alignment with or changes to their existing practices and the preservation of authorial agency (§\ref{sec:6.3}). Lastly, we present findings on participants’ expectations for using Constella when writing the entire full-length story, suggested improvements for supporting varied writing processes, and system limitations such as generation quality and insufficient support for narrative advancement (§\ref{sec:6.4}).


\begin{figure}[H]
    \centering
    \includegraphics[width=\textwidth]{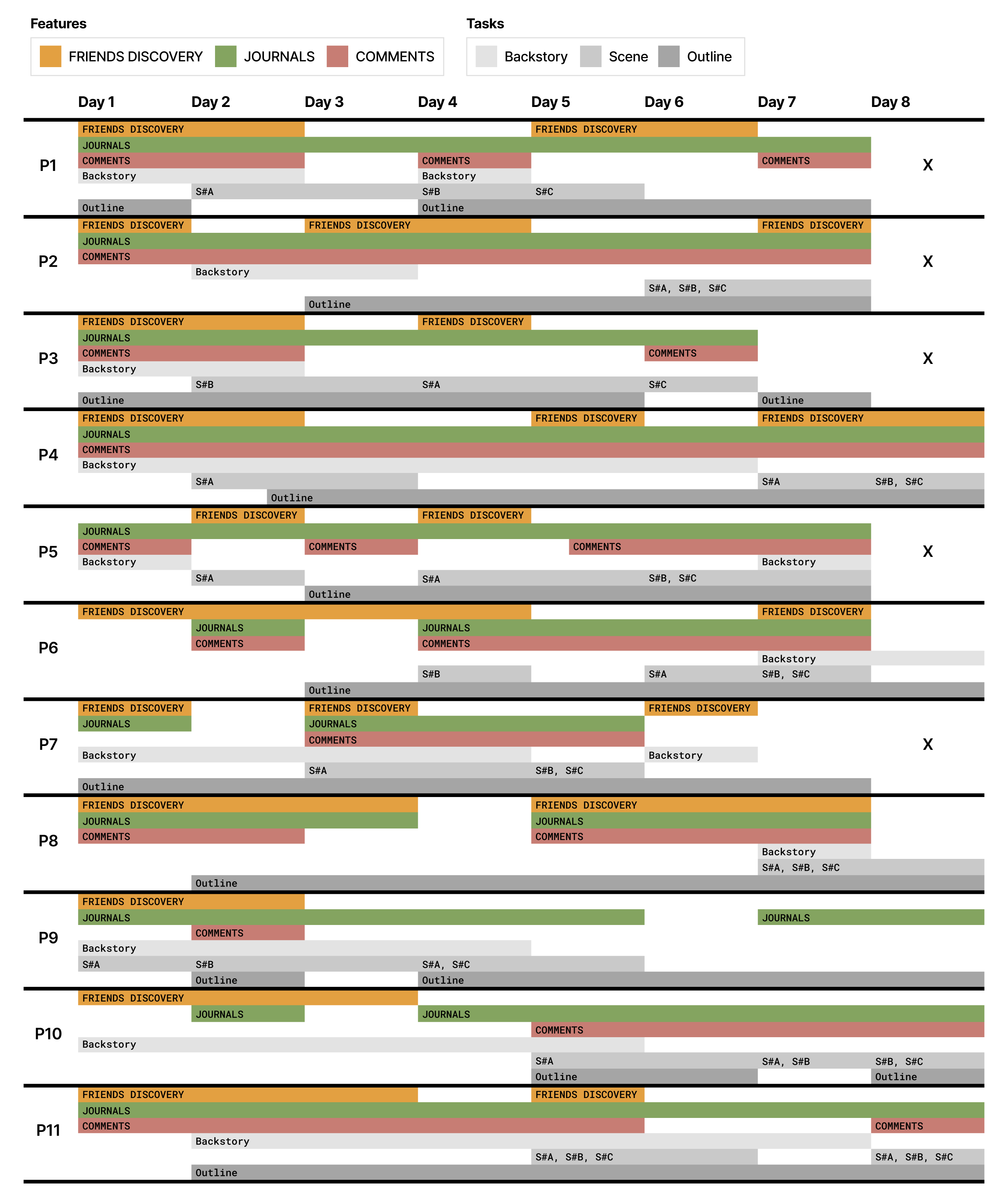}
    \caption{Timeline of feature usage and task engagement across participants over the 7–8 day study period. Colored bars indicate usage of Constella’s AI features: FRIENDS DISCOVERY (yellow gold), JOURNALS (olive green), and COMMENTS (red clay). Grayscale bars represent user engagement with their study tasks: Backstory, Scenes (S\#), and Outline. The ‘X’ denotes participants who completed the study by Day 7, indicating no further engagement on Day 8.}
    \Description{A horizontal timeline chart with rows for participants P1–P11 and columns for Days 1–8. Each row contains stacked colored and grayscale bars spanning different days.}
    \label{fig:timeline}
\end{figure}

\subsection{Overview}
\label{sec:6.1}
    \subsubsection{System Logs}
    \label{sec:6.1.1}
    \autoref{fig:timeline} provides a timeline of the tasks participants completed and Constella’s AI features they engaged with over a 7–8 day period. As illustrated in \autoref{fig:timeline}, participants approached the tasks in varying orders and durations, aligning with their individual creative processes. Many also leveraged Constella beyond the scope of the assigned tasks, tailoring its features to meet their personal needs. This variability highlights both the flexibility of our task design and the adaptability of Constella, demonstrating how they both supported personalized and self-directed creative processes. We now present a deeper analysis of \autoref{fig:timeline}, along with our quantitative analysis results.

    \paragraph*{\textbf{FRIENDS DISCOVERY}} Some writers (e.g., P9, P10) used this feature exclusively during the early stages of their process, suggesting a preference for establishing their entire character cast before delving into story composition (see yellow gold-colored bars in \autoref{fig:timeline}). Others (e.g., P4, P6) used this feature throughout their writing process, indicating its utility in continuously introducing new characters as the narrative evolved. In total, 41 out of 89 characters and on average 3.73 characters ($SD=1.95$) out of 8.09 characters ($SD=1.81$) featured in their work were initially created through FRIENDS DISCOVERY. These 41 characters spanned a wide range of identities, including a wizard, psychopath, astronomer, security guard, and aunt. Examples of prompts used by participants include "Hosu’s co-worker" (P3), "At first, they had a hostile relationship, but gradually, they came to understand and love each other" (P6), and "A relationship bound by secrets that must never be revealed" (P11). 

    \begin{table}[H]
\centering
\caption{Results of the editing metrics for profiles, journals, and comments computed from the system logs. Edit counts represent the number of revision operations performed by the user. For word retention, 0 means no words were retained, and 1 means all words were retained. For word insertion, 0 means no words were inserted, and 1 means all words were inserted. Values are reported as \textit{mean (standard deviation)} and are calculated per profile, journal entry, or comment.}

\begin{tabular}{lcccc}
\toprule
& \multicolumn{2}{c}{\textbf{Profile}} & \textbf{Journal} & \textbf{Comments} \\
\cmidrule(lr){2-3}
\textbf{Metric} & System & Manual & System & System \\
\midrule

Edit count &
7.85 (7.62) & 13.35 (13.87) & 0.04 (0.30) & 0.05 (0.24) \\

Word retention rate &
0.77 (0.28) & 0.75 (0.32) & 0.99 (0.06) & 0.97 (0.15) \\

Word insertion rate &
0.37 (0.29) & 0.68 (0.34) & 0.00 (0.03) & 0.02 (0.12) \\
\bottomrule
\end{tabular}
   \label{tab:editing-metrics}
\end{table}

    To more clearly examine how writers refined characters seeded by FRIENDS DISCOVERY, we analyzed text edits made to their profiles while comparing them with characters that participants introduced manually (\autoref{tab:editing-metrics}). Overall, we find that generated characters are not simply accepted verbatim. Writers still revise them, with an average of 7.85 edits per profile ($SD=7.62$). A major proportion of the generated wording is preserved (word retention rate $M=0.77$, $SD=0.28$) and selectively expanded (word insertion rate $M=0.37$, $SD=0.29$). In contrast, manually created characters exhibit more intensive revision, with more edit counts ($M=13.35$, $SD=13.87$), comparable content retention (word retention rate $M=0.75$, $SD=0.32$), but noticeably heavier expansion (word insertion rate $M=0.68$, $SD=0.34$).

\begin{figure} [H]
  \includegraphics[width=0.7\textwidth]{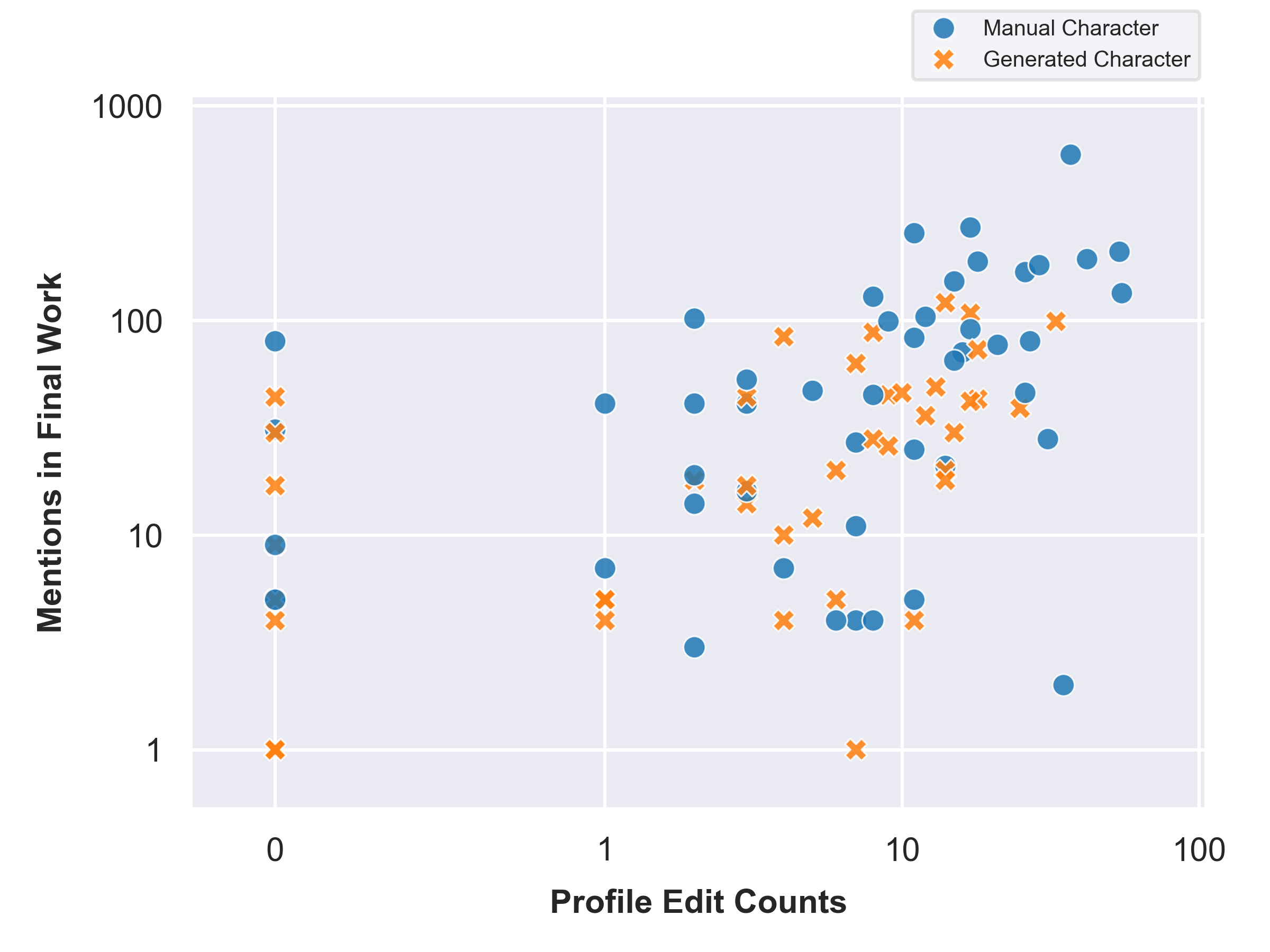}
  \caption{A scatter plot showing the relationship between per-character profile edit counts and mentions in the participants’ final work. Each point represents a character (blue = manually created, orange = generated). The x-axis shows the profile edit counts, and the y-axis shows the number of name mentions in the final work. We used a symmetric logarithmic (symlog) scale \cite{Hunter2007,webber2012bi}, also employed in prior research \cite{10.1145/3297663.3310302,10.1145/3196494.3196516}, to include zero values in the visualization while compressing the wide range of edit and mention counts so that both small and large values remain visually comparable.} 
  \label{fig:scatter-plot}
   \Description{Points in the scatter plot show an overall upward trend, with higher edit counts generally corresponding to more mentions. This positive association appears for both manual and generated characters, although some manual-character points extend to higher mention and edit levels.}
\end{figure}

    This distinction in refinement behavior, however, must also be understood in relation to the roles these characters play in the narrative. Turning to writers’ final work, we found that manually created characters were mentioned more frequently ($M=78.02$, $SD=104.63$) than generated characters ($M=32.49$, $SD=31.52$). In addition, using Spearman’s correlation (see \autoref{fig:scatter-plot} for the scatter plot), we found that across all characters, the number of profile edits was strongly associated with their mentions in the final work ($r_s$(87) = .58, $p < .001$). This pattern held for both manually created characters ($r_s$(46) = .55, $p < .001$) and generated characters ($r_s$(39) = .59, $p < .001$). 

    Together, these results point to two distinct patterns of refinement: manually created characters, often central in their stories, tended to undergo an expansion-driven process where initial profiles served as seeds for heavy elaboration. Generated characters, often less central, likely served as strong initial scaffolds that writers selectively refined. Importantly, the strong association between editing counts and final mentions suggests that, within each origin, writers ultimately invest more refinement in the characters that play larger roles in their final work.

    \paragraph*{\textbf{JOURNALS}} This feature was the most consistently used AI feature (see olive green-colored bars in \autoref{fig:timeline}). Although initially assigned for Scene Type A, several participants (e.g., P1, P2, P5) employed this feature daily. They expanded its use to tasks such as backstory writing and selectively applied it across other tasks as well. By the end of the study, participants’ Constella accounts contained a total of 533 individual journal entries created by characters featured in their work. Of these, 441 were created using the feature’s option that allows multiple characters to produce an entry on the same theme. On average, participants’ accounts contained 48.45 entries ($SD=42.43$). These entries covered a range of themes, including characters’ backgrounds, daily lives, thoughts about each other, and situations closely related to their stories. Specific journal themes included "A lucky day" (P1), "Chaeyeon was caught by Hyunsu while trying to steal chemicals in the science lab and failed. What do you think?" (P2), and "Before Choi Ji-ah is eaten by the monster, Kim Chul-soo rescues her. He also saves Choi Ji-ah’s parents. Because of this, Kim Chul-soo becomes a benefactor to Choi Ji-ah’s family. At this moment, Choi Ji-ah awakens as a hero. She then decides to help Kim Chul-soo seek revenge." (P10).

    To better understand how writers actually made use of the generated journal entries, we first analyzed their editing behavior across all generated entries (\autoref{tab:editing-metrics}). We found that they were rarely revised within the tool, having only an average of 0.04 edits per entry ($SD=0.30$). Text-level edit metrics likewise indicate minimal rewriting, with extremely high generated word retention ($M=0.99$, $SD=0.06$) and almost no new text added (word insertion rate $M=0.00$, $SD=0.03$). One interpretation is that writers may have treated generated journals as low-stakes scaffolds for thinking and planning, rather than as direct story text that demands continuous revision.

    To further analyze whether this interpretation holds true, we examined sentence-level reuse between generated journal entries and participants’ final work, combining exact sentence matches with a Jaccard-based similarity measure. As a result, we found that a tiny fraction of journal sentences was reused in writers’ final work. Across all participants, only 27 out of the 11,115 journal sentences overlap with any sentence in the final work, meaning that roughly one out of every 400 journal sentences is reused. While this provides quantitative support for our interpretation, these metrics cannot fully capture semantically transformed reuse or more subtle forms of influence. Therefore, we complement this with our qualitative analysis to provide a more complete picture of how writers draw on journal entries, which we elaborate on in §\ref{sec:6.2.2}.
     
    \paragraph*{\textbf{COMMENTS}} This feature was used more intermittently compared to JOURNALS (see red clay-colored bars in \autoref{fig:timeline}). This suggests that participants employed the feature for relatively more focused or limited purposes. Among those that were created using characters featured in their work, a total of 175 comment threads with an average of 15.91 threads per participant ($SD=13.48$) were left in their Constella accounts upon completing the study. 
    
    As with JOURNALS, we analyzed how generated comments were edited to further understand writers’ usage patterns (\autoref{tab:editing-metrics}). Comments were also edited rarely within the tool, with an average of 0.05 edits ($SD=0.24$) per comment. Writers largely kept the generated wording (word retention rate $M=0.97$, $SD=0.15$) and added almost no new text (word insertion rate $M=0.02$, $SD=0.12$). Similar to JOURNALS, generated comments were mostly left as-is, suggesting that they may have also been used as a side channel for exploration rather than a place to develop polished narrative text.

    To probe this interpretation, we examined how much of the participants’ final work could be traced back to generated comments by analyzing sentence-level reuse, combining exact sentence matches with a Jaccard-based similarity measure. Across all participants, only 7 out of the entire 2,135 comment sentences overlap with any sentence in the final work, meaning that roughly one out of every 300 comment sentences is reused. While this indicates that comments are rarely reused verbatim or near-verbatim in the main narrative, it does not directly mean they were inconsequential. As with JOURNALS, we show qualitative cases that reveal how COMMENTS informed writers in §\ref{sec:6.2.3}.

  \subsubsection{Creativity Support Index}
    \label{sec:6.1.2}
\begin{table}[ht]
    \renewcommand{\arraystretch}{1.3}
    \caption{Creativity Support Index results. Participants rated Enjoyment ($M=17.45$) as Constella’s strongest factor, followed by Results Worth Effort ($M=16.73$), Expressiveness ($M=16.55$), Exploration ($M=16.00$), Collaboration ($M=15.45$), and Immersion ($M=14.36$). Although we did not include any collaboration tasks between the participants, we nevertheless administered the complete CSI survey, including the Collaboration-related questions. This was to use a standardized survey metric by not changing the original questions and calculation methods as recommended in the original paper that introduced the CSI \cite{10.1145/2617588}, a practice also observed in other studies that evaluate individual tool usage \cite{10.1145/3450741.3465253, 10.1145/3613904.3642105}. Moreover, while the Avg. Factor Score for Collaboration is less informative in our single-user setting, the extremely low Avg. Factor Count for Collaboration confirms that participants regarded collaborating with other people as unimportant for their given tasks, resulting in this factor carrying minimal weight (Avg. Weighted Factor Score) in the final CSI score.}
    
    \label{table:csi-results}
    \resizebox{\textwidth}{!}{%
      \begin{tabular}{@{}T{3cm} T{3.4cm} T{3.7cm} T{5cm}@{}}
      \toprule
      \textbf{Scale} &
      \textbf{Avg.\ Factor Score (SD)} &
      \textbf{Avg.\ Factor Count (SD)} &
      \textbf{Avg.\ Weighted Factor Score (SD)} \\
      \midrule
      Results Worth Effort   & 16.73 (3.10) & 3.45 (1.37) & 58.36 (26.67) \\
      Exploration            & 16.00 (2.61) & 4.55 (0.69) & 72.82 (16.98) \\
      Collaboration          & 15.45 (3.42) & 0.36 (0.92) &  7.27 (18.49) \\
      Immersion              & 14.36 (2.94) & 1.73 (1.10) & 23.27 (15.76) \\
      Expressiveness         & 16.55 (2.81) & 3.09 (1.22) & 50.45 (21.17) \\
      Enjoyment              & 17.45 (2.62) & 1.82 (0.60) & 32.00 (11.25) \\

      \midrule
      \textbf{Overall CSI Score} & \textbf{81.39 (12.02)} & & \\[0.2em]
      \bottomrule
      \end{tabular}}%
\end{table}
  
    Constella received an average CSI score of 81.39 out of 100 ($SD=12.02$) (\autoref{table:csi-results}). Among the six CSI factors, "Enjoyment" received the highest individual score ($M=17.45$). Further analysis of our qualitative data shows that a key contributor to this score was the social media metaphor. Beyond helping participants (P2, P3, P4, P6, P8, P10, P11) intuitively understand tool usage and thereby feel more at ease, the social media metaphor was regarded by writers (P2, P4, P6, P10) as helping to foster a playful environment. As P2 noted, \textit{"The format of social media, such as showing when someone is online and including comments, made it feel playful to me."} In addition, the specific interaction concepts of the social media metaphor also led participants to playfully envision fiction-writing techniques and narrative possibilities grounded in its unique affordances. For example, P10 speculated on using the asynchronous nature of COMMENTS to create dramatic irony, explaining, \textit{"Until a reply is made, [the diary writer] doesn’t know the content of the comment, right? I think using that kind of setup could be really fun. [...] like when the reader and the protagonist know something, but the villain doesn’t." }

    Conversely, "Immersion" was the lowest-scoring CSI factor ($M=14.36$). One explanation provided by our participants was the need to switch back and forth between their word processors and Constella, which disrupted their focus. We initially chose not to include a built-in word processing feature out of respect for writers’ existing tool preferences. However, based on our findings, we touch upon why integrating such functionality may be useful in §\ref{sec:6.4.2}.

\subsection{Use Cases for Interconnected Character Creation}
\label{sec:6.2}
We report several use cases for each key feature in Constella to illustrate how these features support storywriters’ interconnected character creation process.

    \subsubsection{FRIENDS DISCOVERY}
\label{sec:6.2.1}

\noindent
\begin{shaded}
\textbf{Contextualized Discoveries:} Writers were able to create diverse characters linked to the central figure in contexts they desired. The feature also enabled them to uncover unexpected yet meaningful connection points. 
\end{shaded}

    All participants either explicitly stated or generally agreed that they were able to discover various characters in contexts they desired. For example, P10 described using this feature by first writing a detailed backstory in the protagonist’s profile. Then, by entering the simple relationship phrase "revenge" in the input field of FRIENDS DISCOVERY, he discovered three characters that not only aligned with but also expanded his initial conception of the antagonists.
    
    Additionally, participants (P3, P5, P6, P8) expressed astonishment when the tool surfaced connections they had not initially considered. P6 shared a case in which the tool suggested a character associated with the element of "swimming." This unexpected detail turned out to be crucial for her story, offering a natural and meaningful way for the protagonist to meet the sub-protagonist, the suggested character, for the first time. 
    \begin{quote}
        \textit{"I had only briefly mentioned swimming as part of the protagonist’s daily routine, which I included as a small detail while writing [the protagonist’s profile]. But when it suddenly emerged [in the suggested character], I realized that the system had considered even these minor details and made connections."} (P6) 
    \end{quote}
    Finally, participants (P5, P11) expressed high satisfaction with the recommended characters, noting that they were stylistically consistent with their central characters. This coherence allowed new characters to be integrated smoothly, without disrupting the story’s overall mood or tone.
    \begin{quote}
        \textit{"When I wanted a character with a vibe similar to Character A  [...] it would provide recommendations for characters that matched the similar vibe I wanted but offered slightly different variations."} (P5)
   \end{quote}

    \vspace{1em}
     \begin{shaded}
    \noindent
    \textbf{Story World and Narrative Expansion:} Generated characters often transcended their initially envisioned roles and contributed to the narrative in new and unforeseen directions.  
    \end{shaded}
    
    Participants used FRIENDS DISCOVERY to identify new characters that aligned with their initial ideas, which they then further developed based on their narrative needs. However, there were also instances where participants (P1, P3, P4, P5, P7, P8, P11) derived entirely new ideas from the recommended characters, which in turn sparked the creation of new storylines. For example, P4 shared that the subplot she developed around a supporting character, initially generated through Constella, felt even more compelling than the protagonist’s storyline she had initially conceived. P4 continued:
    \begin{quote}
        \textit{"Originally, creating main characters requires a lot of thought, so paying attention to supporting characters or secondary events usually happens later in the writing process. But this method feels new and helpful. As I included the recommended supporting characters in my story and they merged with my protagonist’s narrative arc, I was able to create a whole new story."} (P4)
    \end{quote}
    Some participants (P3, P8, P11) even adopted the experimental approach of generating entire communities of related characters early in their ideation process, then built their narrative around these characters. For instance, P11 described how she built two distinct story worlds, populated with a village-like network of interconnected figures, each centered around a different character.
    \begin{quote}
        \textit{"In my case, I initially created many supporting characters around the main character, Yoo Ji-an. I kept imagining a world connected to her, and then at some point, I wanted to create a completely different world. [...] By continuing to generate related characters, I ended up with two very different worlds inhabited by distinct characters."} (P11) 
    \end{quote}
    
    \vspace{1em}
    \begin{shaded}
    \noindent
    \textbf{Stockpile Resources:} Generated characters often served as stockpile resources, flexible components ready to be divided, combined, or set aside for use in unexpected situations.  
    \end{shaded}
 
    While generated characters were often used as individual entities, some participants (P1, P2, P3, P10, P11) selectively incorporated elements from the generated characters into existing ones or synthesized them into their ideal characters. P1 noted that this was possible because Constella provided multiple character recommendations at once.
    \begin{quote}
        \textit{"Since about three characters are created, I was able to combine various characteristics from them to shape the character I wanted, which was definitely helpful."} (P1) 
    \end{quote}
    Furthermore, P2 saved suggested characters even if they were not immediately needed in her story, in case she might need them in the future. 
    \begin{quote}
        \textit{"By creating and saving characters, I could bring them back later as cameos when the story hit a dead end, allowing the narrative to take a different direction."} (P2)
    \end{quote}

    \subsubsection{JOURNALS}    
     \noindent
    \begin{shaded}
    \textbf{Introspective Construction:} JOURNALS helped writers delve into the inner worlds of their characters, including those who might have otherwise remained sidelined or underdeveloped.  
    \end{shaded}
    
    All participants mentioned that JOURNALS helped delve into their characters’ psychology, thoughts, and emotions. In addition, some writers (P2, P4, P8, P9, P11) noted that the feature also facilitated the development of the supporting cast. For instance, P8 even wrote a scene from the villain’s perspective for his task, explaining that this was possible because the feature helped him reflect deeply on the character’s emotions and ideology. P8 further explained that using the feature led him to humanize the antagonist, endowing the villain with human qualities that made it difficult for him to view the villain solely with contempt. P9 similarly noted that JOURNALS was useful for concretizing the backgrounds of characters that might otherwise have been used in a throwaway manner. 
    \begin{quote}
        \textit{"I think there are characters who inevitably end up being used in a disposable way for the protagonist. While working on the detailed background of such characters [through JOURNALS], I found that I could effectively use these characters to shape scenes or plots that could appear later on."} (P9)
   \end{quote}
    JOURNALS was also perceived to help writers (P4, P10, P11) deeply immerse themselves in several characters simultaneously. For instance, P11 noted that this helped her distribute attention and emotional investment more evenly across characters.
    \begin{quote}
        \textit{"When a writer keeps writing the dialogue or situation for two characters, they might end up becoming more immersed in one character or fail to fully understand the perspective of the other. [...] But by creating [journal entries] first and then reading through both perspectives, I found it helped me approach the process more objectively and distribute [my] emotions more evenly between the two characters."} (P11)
    \end{quote}
     \vspace{1em}
     \noindent
        \begin{shaded}
    \textbf{Relational Construction:} JOURNALS helped writers create characters in relation to one another, either by allowing them to view one character through others’ perspectives or by exploring how multiple characters respond differently to shared themes.
    \end{shaded}
    
    By viewing \textit{Character B} from \textit{Character A}’s perspective through the journal entries, participants (P2, P7, P9) were able to further develop \textit{Character B}. For example, P7 mentioned that viewing "Jane" through the eyes of "Taesung" helped him imagine Jane’s identity in greater depth. Similarly, P2 shared her experience about a journal entry as follows:
    \begin{quote}
        \textit{"As I started to look at the new situation of the security guard, a supporting character, through the [main character’s] eyes, I found myself wondering, what might have happened to the man? The wind blew a lot, and it rained. Were there any difficult or troublesome complaints he had to deal with?"} (P2) 
    \end{quote}
    Participants (P1, P2, P4, P5, P7, P8, P9, P11) also stated that having two or more characters generate journal entries based on a shared theme helped them distinguish or create similarities between characters. Specifically, P1 noted that observing diverse thoughts across several characters aided in the creation of each of their dialogue, narrations, and non-verbal actions. P4 also explained that comparing journal entries helped in plot creation. 
    \begin{quote}
        \textit{"When I asked the same question to multiple characters, I noticed differences between them within the journals, and by comparing those differences, I was able to create new scenes and storylines."} (P4)
     \end{quote}
     \vspace{1em}
    \noindent
    \begin{shaded}
    \textbf{Mise-en-Scène\footnotemark{} of the Mind:} The visual elements depicted in the journal entries were used to express the mindscapes within and between characters.
    \end{shaded}
    \footnotetext{Mise-en-scène refers to the visual elements within a frame,
    including lighting, color, and the arrangement of objects \cite{kuhn2012dictionary}.}

    As JOURNALS often generated descriptive details of the characters’ surrounding environments, participants (P1, P2, P5, P9, P11) drew inspiration from the locations, landscapes, and weather described in the entries. These spatial and environmental cues served not only as narrative settings but also as symbolic reflections of characters’ internal states. For example, P11 used such imagery to internalize a character’s emotional condition and developed scenes based on her interpretive reflections:
    \begin{quote}
        \textit{"Descriptions of Jian’s feelings [in her journal entry], like wanting to return to a green forest but being trapped in a gray city, these types of phrases weren’t directly used as dialogue. Instead, they evoked certain imagery or moods for me. When I absorbed these, some kind of chemical reaction happened within me, and I was able to create scenes."} (P11)
    \end{quote}
    Furthermore, P9 drew on the image of "sunlight" from a journal entry not to suggest emotional warmth, but to emphasize its absence, showing how even a bright day cannot dispel the chill between the characters.
     \begin{quote}
        \textit{"I think I spent a lot of time thinking about how to use ‘sunlight’ effectively. As I looked through the journal, I realized [the protagonist] had never felt warmth from Sunwoo. Their meetings had always been cold, happening at night. So I wanted to show how even that daytime setting gets overtaken by the same nighttime mood."} (P9) 
    \end{quote}

    \label{sec:6.2.2}
    \subsubsection{COMMENTS}
    \label{sec:6.2.3}
     \noindent
        \begin{shaded}
    \textbf{Extra Layers:} COMMENTS was mainly used as an extension of the generated journal entries, allowing writers to further explore character relationships through additional responses. 
    \end{shaded}

    Some participants found COMMENTS to be more effective than JOURNALS for understanding the psychological interplay between characters (P2) or for establishing character relationships (P7). While JOURNALS supported the development of individual emotional arcs through monologue-like reflections, COMMENTS was more effective for exploring interpersonal emotional dynamics. This was because the reply-like format of COMMENTS, even when used in brief, one-off responses, mimicked communicative interactions.

    In a similar vein, participants (P2, P4, P6, P7, P8, P10, P11) mentioned that observing character responses through COMMENTS helped them further elaborate on relational subtleties. For example, P4 used COMMENTS to detect subtle cues of ill intent, which informed her decisions about whether those characters could plausibly collaborate as allies. In contrast, P10 generated comments to reinforce existing dynamics rather than introduce new shifts in relationships.
    \begin{quote}
        \textit{"I noticed through the comments that, even after such humiliation and having his Achilles tendon severed, [Lee Min-jun] doesn’t give up. That made me want to crush [Lee Min-jun] even more. I think those impressions ended up being reflected in the scenes."} (P10)
    \end{quote}
    
\subsection{Impact on Writers and Their Writing Process}
\label{sec:6.3}
    \subsubsection{Alignment and Changes to Writers’ Existing Practices}
    \label{sec:6.3.1}
      \paragraph*{\textbf{Alignment}} All participants (N=11) agreed that FRIENDS DISCOVERY aligned well with their established practices. They explained that this feature mirrored their usual approach of conceptualizing new characters in relation to existing ones. JOURNALS also aligned well with most writers’ existing practices (N=9). P4 specifically mentioned that this feature was similar to her established method of considering how different characters would each react to a shared situation. COMMENTS received mixed feedback in terms of effective practice alignment. While participants who expressed positive remarks (N=6) mentioned the usefulness of this feature to explore character relationships, others who gave negative feedback (N=5) pointed out that COMMENTS felt unnecessary or inefficient compared to JOURNALS. This was because JOURNALS could be utilized to serve similar purposes as COMMENTS, but conveyed more extensive details due to their longer format.

    \paragraph*{\textbf{Changes}} Constella also introduced notable shifts in participants’ existing writing practices. Writers (P2, P3, P6, P8, P11) who typically initiated story ideation with conflicts or a logline reported that our tool nudged them to begin with character development instead. Specifically, they mentioned being immediately drawn to the Create Character button (see \autoref{fig:fig4}) in the user interface (P2) and wanting to first create numerous characters to get used to the tool (P3, P8, P11). These characters then gradually became interesting enough to build their plot around (P3) and became a space for personal and enjoyable experimentation as writers explored what new stories could emerge by using them (P8, P11).
    
    Participants also appreciated how Constella shifted their focus towards deeper engagement with a broader range of characters, including those they might not have even considered at all, at an earlier stage in their writing process. Furthermore, some writers (P8, P10) became curious about exploring the perspectives of characters not directly involved in the current story situation. P10, in particular, enjoyed using this newfound ability to speculate on how other unassociated characters might react alongside those more centrally involved, \textit{"since the rumor can spread to others."} 
    
    Lastly, Constella offered valuable insights for participants to refine their creative methods in the future. P3 came to appreciate the value of having a diverse pool of characters that could be flexibly integrated into any scenario. Other writers (P8, P9) expressed a desire to change their usual approach of organizing character attributes and relationships, inspired by the structure and layout of the Profile feature.
    
    \subsubsection{Preserving Authorial Agency}
    \label{sec:6.3.2}
    Participants perceived Constella to play a complementary role in their writing process, with some (P7, P8, P10, P11) referring to it as an \textit{"assistant writer"}. Our analysis suggests such perceptions were shaped by specific design aspects of Constella that supported writers in maintaining their creative voice within human-AI collaboration. Below, we examine these design elements in greater detail. 

    \paragraph*{\textbf{Constraining AI Involvement}} Since FRIENDS DISCOVERY generates characters only in relation to existing ones, the responsibility of creating profiles for the initial characters, typically the main protagonists, was placed on the writers. Participants (P7, P11) noted that this feature resembled their co-writing dynamics, where they delegate secondary character creation to collaborators. They also mentioned that creating descriptions for the main character themselves helped ensure a greater sense of ownership of their work (P10, P11). Furthermore, as the tool only provides core character attributes, some writers (P3, P11) noted that this created room for them to build on the suggested characters with their own ideas.
    \begin{quote}
    \textit{"If the main characters were also created by the tool, I think it could be like someone addicted to gambling, just staring blankly at the app, completely absorbed. But since only supporting characters are generated and there’s a certain limit, it made me snap out and think, wait a minute, I need to create this part myself."} (P11)
    \end{quote}

    In addition, some writers (P1, P3, P8, P11) noted that Constella demanded a high level of initiative. As the tool is tightly scoped to provide AI support for character creation, P3 mentioned that having a clear direction for the plot, which he regarded as his responsibility and not AI’s, was needed to use this tool meaningfully. In a similar vein, as FRIENDS DISCOVERY and JOURNALS require writers to craft their own prompts without any AI-generated prompt suggestions, P1 felt that she needed to think through what she was aiming for prior to feature usage. This led her to view Constella as an extension of her expressive intent rather than a replacement for it.

     \begin{quote}
    \textit{"If I want to find a character B that contrasts with character A [through FRIENDS DISCOVERY], I have to write in how I want them to contrast, right? That part requires me to think it through, and the character that gets created is based on the ideas I’ve already formed in my head."} (P1)
    \end{quote}

    Thus, while there were instances where writers pursued new narrative directions inspired by initial experiments with Constella (as mentioned in §\ref{sec:6.3} – Changes), these findings show that writers still had to exert ongoing responsibility for key aspects of the creative process.
    
    \paragraph*{\textbf{Intermediary Materials}} We designed JOURNALS and COMMENTS to generate text in formats that resemble diary entries and responses to these entries. Participants (P1, P3, P9, P10) mentioned that they were initially surprised because they had only heard of AI tools that generate entire outlines or stories in output-ready formats. However, for our tool, writers (P1, P2, P7, P8, P9, P10) pointed out the importance of their own role in consciously extracting meaning from the generated results. P8, for instance, noted that since JOURNALS did not simply give an answer, the ability to catch and further develop the content according to his own criteria was an important skill needed when using the tool. He explains, \textit{"I don’t believe a person can suddenly do something [with the outputs] that lies completely outside their own taste or abilities. If you try to use things beyond what that person actually has, the story is bound to fall apart."} Adding to this, P10 described how the tool differs from other AI systems, as he remained the one in control when shaping the scenes.

    \begin{quote}
      \textit{"Usually, when people think of AI, they assume it just generates stories automatically when given a prompt.  But in that case, AI is the one leading the process. In contrast, when I provide the input, this tool creates  journals and other things, and through these journals, I create my scenes."} (P10)
    \end{quote}
 
Thus, these generations seemed to sit between the writer and their final work as intermediary materials, ready for further adaptation and meaning-making (§\ref{sec:6.2.2}) before blending into the work.

\subsection{Future Usages, Suggested Improvements, and Limitations}
\label{sec:6.4}
    \subsubsection{Potential while Writing the Entire Full-Length Story}
    \label{sec:6.4.1}
    Participants suggested that Constella would be utilized even more extensively when writing the main text of the story, rather than summarized outlines. When engaged in this process, writers (P1, P3, P7, P8, P10) noted they would use the tool to delve even deeper into characters’ backstories and thoughts, concretize their relationships, and create more supporting characters. P8 also shared that using Constella to develop identities for background characters, instead of simply treating them as extras, would even help imagine their visual appearances during the illustration process.

    In addition, participants (P1, P3, P8) mentioned that Constella would be more beneficial when used for large-scale stories that require expansive worldbuilding and complex relationships, such as serialized works and multiple stories that share the same fictional world. Considering such situations that require vast numbers of interconnected characters, updating FRIENDS DISCOVERY to generate characters associated with multiple individuals (P2), including character relationship maps (P1, P10, P11), adding a tagging feature (e.g., @name) to easily change and reference character names (P10, P11), and organizing characters into groups (P3, P6, P8) could further help writers expand and manage their character casts.

    \subsubsection{Additional Support for Writers’ Linear, Non-linear, and Iterative Processes}
     \label{sec:6.4.2}
    Participants (P2, P11) requested a feature that allows characters to recall previously generated journal entries and comment threads to explore character relationships that evolve. Additionally, a version control \cite{10.1145/3555756} feature to track and recover changes in character profiles was suggested (P3, P4, P7, P10). Specifically, P4 expressed interest in exploring characters non-linearly by aligning profile versions with different points along the story’s timeline. While writers often created journals for a specific purpose, they also derived new meanings when reviewing entries in sequence or looking back on them later (P3, P4, P9, P11). As a result, a feature that allows users to reorganize journal entries into new categories or story timelines was also requested (P3, P4). To better support the iterative process of editing character profiles, participants (P3, P7) suggested integrating a word processor into Constella, allowing them to seamlessly make changes while writing their stories within a single interface.

    \subsubsection{Generation Quality and Intent Elaboration}
     \label{sec:6.4.3}
     Some writers (P5, P8) mentioned that JOURNALS and COMMENTS were sometimes overly constrained by character attributes, leading to a lack of novelty. For instance, P8 mentioned that the system tended to echo character descriptions too literally; if a character was described as disliking someone, that sentiment would appear in the generated text regardless of the overall context. 
     
    In addition, a common critique was that the generated results felt somewhat mundane or stereotypical (P1, P2, P3, P5, P6, P9), and the characters were depicted as excessively kind (P1, P2, P5, P6, P11). However, many of these writers (P1, P2, P9, P11) also mentioned that their perceptions shifted to varying degrees after creating more detailed character profiles or prompts. They also believed that additional effort in doing so could have yielded better results (P1, P3, P5, P6, P11). That said, participants found the process of creating detailed character profiles challenging. Specifically, writers (P2, P3, P8, P11) mentioned a trade-off between the effort required in detailing character profiles and the benefit of obtaining consistent, high-quality generations. For example, writers (P2, P3, P8) did not input all character details (e.g., nationality, manner of speech) as they typically only write down key traits when working alone. They also mentioned that adding descriptions to achieve more aligned results felt frustrating since they had already mentally envisioned those characteristics. Continuously revising character profiles to incorporate updated information from the storyline also required considerable effort (P8, P11). We discuss potential methods to alleviate these issues in §\ref{sec:7.5} and §\ref{sec:8}.

    \subsubsection{Limited Narrative Advancement}
     \label{sec:6.4.4}
     While acknowledging that this was not Constella’s intended purpose, some writers (P1, P5, P7, P11) expressed disappointment that the generated results for COMMENTS and JOURNALS were not presented in forms they found \textit{"directly applicable to the story"} (P7) or had limited ability to \textit{"progress the narrative"} (P11). This was largely because the outputs lacked sufficient action sequences (P5, P7, P11) or were inadequate for simulating in-the-moment conversational interactions mimicking the form of dialogue in stories (P1, P11). Instead, the outputs were focused more on characters’ thoughts, emotions, and reflections on past events. While this approach was perceived as a shortcoming by some, we offer our reflections on how it may have encouraged writers to exercise greater creative judgment in §\ref{sec:7.3}.


\section{Discussion}
\label{sec:7}
In this section, we first identify opportunities for AI interaction design in storywriting. To do so, we begin by reflecting on our system’s design goals through revisiting patterns of user appropriation (§\ref{sec:7.1}) and exploring how multi-agent interactions can be intentionally designed to encourage balanced attention across the character cast (§\ref{sec:7.2}). We then present broader design implications for AI-based writing support tools that value authorial agency by highlighting the value of deliberate constraints and intermediary materials (§\ref{sec:7.3}). We also discuss the opportunities and challenges of employing metaphors in AI-based CSTs (§\ref{sec:7.4}). Finally, we address the challenge of intent alignment, advocating for AI assistance that reduces cognitive load without diminishing opportunities for user reflection (§\ref{sec:7.5}).

\subsection{Reflecting on the Design Goals}
\label{sec:7.1}
We reflect on our system design goals outlined in §\ref{sec:3.6}, assessing the extent to which they have been achieved through our user study. We also use participants’ acts of appropriation \cite{10.1145/3586183.3606831} and misuse \cite{10.1145/3706598.3714068} uncovered through the study as resources to identify design opportunities for future AI-based storywriting support tools. 

\paragraph*{\textbf{DG 1: Suggest diverse characters related to existing ones}} FRIENDS DISCOVERY allows writers to discover three new characters in relation to a given character by specifying their relationship. In the study, all participants reported that our tool helped them discover various characters aligned with their desired relational contexts (§\ref{sec:6.2.1}). As the tool offered multiple variations, we also observed that several writers (N=4) engaged in a creative process of remixing suggested characters by combining several of their attributes. This emergent use highlights a potential area for improvement: generating new character variations based on remixed attributes. For instance, a writer might select the element of a "magician" from one suggested character and the element of "sleepwalking" from another. The system could then generate further character variations based on their combinations while still aligning with the original relationship prompt. This approach could potentially expand the diversity of characters writers can explore while preserving their relational ties to existing ones. 

\paragraph*{\textbf{DG 2: Support the comparison and investigation of in-depth character identities}} JOURNALS provide writers with characters’ introspective reactions toward a shared theme. All participants in our study regarded the feature as a valuable aid for delving into their characters’ minds and personal viewpoints (§\ref{sec:6.2.2}). In addition, writers (N=8) mentioned that having multiple characters reflect on the same theme helped them better understand and articulate contrasts and similarities between characters. Lastly, we found that several writers (N=5) drew inspiration from the visual and spatial descriptions within the journal entries, treating them as expressive cues of characters’ internal states. In light of these use cases, we propose expanding JOURNALS to support the externalization of characters’ mindscapes through image-generative AI. Such introspective visualizations can leverage the power of ambiguity \cite{10.1145/642611.642653, 10.1145/3706598.3713722} or low fidelity \cite{10.1145/3613904.3642794, 10.1145/3746059.3747605} to facilitate writers’ meaning-making and interpretation while faithfully expressing the emotional texture and mood of each character’s inner world. 

\paragraph*{\textbf{DG 3: Manifest character relationships through exploratory interactions}} COMMENTS enables writers to explore open-ended character reactions toward others’ journal entries. While many writers (N=7) found the feature helpful for surfacing inter-character dynamics (§\ref{sec:6.2.3}), we also observed that JOURNALS was often used for similar purposes by defining themes that prompted characters to reflect on one another (§\ref{sec:6.1.1}, §\ref{sec:6.2.2}). Therefore, to further support relational development in ways that differ from JOURNALS, future work could expand COMMENTS to include non-verbal interactions. For example, characters can intentionally delay their responses, signaling social cues that can be interpreted as hesitation, avoidance, or low commitment \cite{mai2015know, kalman2006pauses}. Characters may also edit or even delete their comments, which may be interpreted as regret \cite{schnitzler2020exploring, lutz2025message} or invite broader interpretation \cite{lutz2025message} when their motivations remain unclear. Accordingly, extending beyond verbal content to include open-ended, nonverbal cues may provide writers with richer materials and inspiration when shaping character relationships.

\subsection{Balancing Attention through Multi-Agent Interactions}
\label{sec:7.2}
We introduced Constella, an LLM-based multi-agent tool that supports storywriters’ interconnected character creation process. Our CSI score (§\ref{sec:6.1.2}), along with qualitative findings that illustrate the diverse ways writers used the tool (§\ref{sec:6.2}), collectively suggest that Constella complemented a range of approaches to character creation. Furthermore, one notable finding was that our tool not only supported but also actively reshaped writers’ existing practices. In particular, Constella enabled writers to broaden the scope of their attention to a greater extent across the character cast (§\ref{sec:6.3.1}).

This complements findings from prior chatbot-based tools for character creation \cite{10.1145/3450741.3465253,10.1145/3613904.3642105}, by showing how AI can support writers in noticing \cite{tsing2015mushroom} and attending to other characters beyond the protagonist. However, this does not imply that all characters receive equal attention; as our quantitative analysis reveals, writers still invest more in their manually created characters with more central roles (§\ref{sec:6.1.1}). Rather, we interpret this as a redistribution of attention, and we coin this emergent phenomenon the \textit{balancing effect}, which allows writers to engage with and further develop characters that would otherwise be overlooked or sidelined. In this section, we examine the underlying features that contributed to the balancing effect, as well as its implications as a design affordance for future storywriting support tools.

Several features of Constella played a critical role in fostering the balancing effect. FRIENDS DISCOVERY surfaced unexpected ties when generating new characters, prompting writers to explore interwoven narratives they had not initially envisioned (§\ref{sec:6.2.1}). We believe this process enabled several generated supporting characters to take on firm and meaningful roles within the narrative, thereby underscoring their potential to shape story dynamics alongside central characters.
Through JOURNALS, writers received a chorus of perspectives from their characters, encouraging them to shift focus more fluidly across the cast (§\ref{sec:6.2.2}). This process helped further deepen and humanize these characters, ultimately reinforcing the notion that these characters, regardless of role, have a story worth telling. We believe this underscores the ability of multi-agent tools to guide writers towards the creation of narratives in which a broader range of characters can potentially contribute more meaningfully to the evolving story. 

Thus, we invite future work to explore how multi-agent systems can be \textit{intentionally designed} to surface underdeveloped perspectives and support more balanced forms of storywriting. For example, multi-agent tools might reimagine a completed scene or even the entire story by exploring how the narrative would change if other supporting characters took the lead. Such generative reframings could be particularly valuable in polyphonic stories, which are characterized by multiple voices, perspectives, and distributed character agency. Given that creative practitioners often adapt their tool choices to intentionally accommodate specific ways of thinking and behaviors \cite{10.1145/3527927.3532870}, and that writers increasingly use AI tools to experiment with their creative processes \cite{10.1145/3698061.3726910, ensemblepark2025}, we believe such LLM-MA tools may hold value not only as creative aids that support well-established practices but also as educational or experimental scaffolds for writers seeking to explore more distributed, multi-perspective, non-traditional storytelling structures. 

While our argument is grounded in prior works that emphasize the value of AI-based CSTs that encourage novel practices \cite{10.1145/3746059.3747605, 10.1145/3706598.3714120}, we also acknowledge that such balancing nudges essentially act as structures that can predetermine their ways of working, thereby exerting subtle forms of power over participants \cite{10.1145/3586183.3606831}. Thus, we believe it will be of utmost importance to transparently communicate the intended purpose of such tools to participants, for instance, during the system tutorial stage as in our study (§\ref{sec:5.2.3}). Doing so can aid creative practitioners when deciding whether and how they should engage with such tools. However, when the changes caused by the tools are unintentional, as in §\ref{sec:6.3.1}, it may be impossible to fully know the tool’s effects prior to user studies and communicate them beforehand.
We thus call for further work to explore alternative interfaces \cite{10.1145/3613904.3641895, glassman2024ai, 10.1145/3613904.3642699, 10.1145/3698061.3726915} that can help users themselves recognize, selectively recalibrate, and resist the tool’s influence as it unfolds.

\subsection{AI Design for Authorial Agency}
\label{sec:7.3}
Our findings indicate that two design aspects of Constella played a critical role in helping writers preserve a sense of agency during human-AI writing: constraining AI’s generative role and offering intermediary forms of text.

\paragraph*{\textbf{Deliberate Constraints}} To align with the idiosyncratic values of writers, existing discourses have emphasized adaptability as a key principle for co-creative AI systems \cite{10.1145/3613904.3642105, 10.1145/3706598.3713146, 10.1145/3532106.3533506}. Yet, might there also be value in applying deliberate constraints on AI within these tools? Our findings show that these constraints, such as not generating the central character and providing only a minimal set of attributes, can prevent writers from delegating creative decisions entirely to the AI system (§\ref{sec:6.3.2}). These imposed boundaries also helped writers avoid over-reliance on AI and encouraged more intentional and thoughtful decisions. 

Echoing research advocating for meaningful limitations in technology \cite{10.1145/2598510.2598522, 10.1145/2207676.2208540, 10.1145/1394445.1394453}, we believe that carefully applied constraints can enable future writing support tools to create room for writers to exercise their creative agency while still drawing on AI’s generative potential. For instance, this could be done by constraining the scope of AI assistance to particular writing stages, as shown by recent work \cite{10.1145/3698061.3726916} that aims to exclusively support prewriting activities. Another example could be a tool that only generates the last sentence of a chapter, prompting writers to build the narrative leading up to that conclusion. Such approaches may also help address what has been described as \textit{"the dearth of the author"} \cite{10.1145/3690712.3690725, kreminski2025endless}, an issue wherein AI systems facilitate large volumes of text with minimal authorial decision-making. Then, how to constrain AI capabilities to help writers maintain authorial control while offering enough meaningful support will be a critical research question to guide the design of future AI-based writing support tools. 

\paragraph*{\textbf{Intermediary Materials}} As mentioned in §\ref{sec:6.4.4}, several writers expected our tool to create action sequences or dialogue in output-ready prose. However, writers also found that the format of JOURNALS and COMMENTS required them to actively engage with the generated materials and think critically about how to use the tool’s outputs in their own creative process (§\ref{sec:6.3.2}). 

Based on these findings, we suggest that a key experiential quality of the intermediary materials produced by our tool is their ability to introduce \textit{slowness} \cite{odom2012slow, hallnas2001slow}. Slowness, when understood as a sign of writing process control \cite{carrera2025add}, indicates that writers are indeed exercising agency over integration decisions. To elaborate, we found that these materials did not propel the narrative forward with polished, ready-made text. Instead, they encouraged the writers to pause, reflect, and develop a more profound understanding of their work or engage in active interpretation before deciding whether and how to incorporate the AI-generated content (§\ref{sec:6.2.2}, §\ref{sec:6.2.3}). 

This resonates with findings from \cite{10.1145/3635636.3656187}, which show that incoherent or grammatically imperfect AI suggestions can provoke reflection and result in more user-generated content. Conversely, our approach contrasts with recent multi-agent NLP research \cite{chen-etal-2024-hollmwood, huot2025agents, yu-etal-2025-multi}, which typically prioritizes speed and automation, given their emphasis on generating complete stories with minimal user intervention.

Along with prior work that aims to preserve \textit{"creative struggle"} in human-AI co-writing \cite{zhou2023creative}, we advocate for AI-generated \textit{"intermediate text"} \cite{10.1145/3635636.3656187}, outputs that are intentionally incomplete or ambiguous, for their value in fostering reflection \cite{kreminski2021reflective} and adaptation during creative tasks. We thus call for future research to further explore alternative forms of AI-based CSTs that do not merely optimize for productivity \cite{10.1145/3698061.3726924} or output-readiness but instead intentionally support slow and interpretive creative processes.

\subsection{Adopting Metaphors for Designing Interactions in AI-based CSTs}
\label{sec:7.4}

Constella was designed with the social media metaphor. In this section, we critically reflect on our own design process to examine the opportunities and challenges that emerged when using metaphors to structure our tool’s interactions, and we broaden this reflection beyond Constella to inform the design of other metaphor-driven AI-based CSTs.

As mentioned in §\ref{sec:4.1}, we adopted the social media metaphor to address our design goals, translating familiar social practices into an interface that intuitively supports character cast interaction. During our design process, working with this metaphor helped us move beyond a common method for interacting with LLMs, the chat-styled interface \cite{10.1145/3450741.3465253,10.1145/3613904.3642105,banterbot2019,10.1145/3706599.3719932}, and instead imagine interactions as acts of shaping a dynamic social space where characters connect, post, and respond to one another. Recent work has similarly used metaphors to design novel interactions, including \textit{Patchview}’s generative dust-and-magnet metaphor for spatially organizing story elements \cite{10.1145/3654777.3676352} and \textit{Toyteller}’s toy-playing metaphor for manipulating symbolic characters to influence narrative generation \cite{10.1145/3706598.3713435}. As such, we believe metaphors can enable more imaginative approaches when designing AI systems for creative activities.

However, one concern persisted throughout the system design process: the potentially divergent nature of social media and storywriting. For instance, because FRIENDS DISCOVERY resembles ‘friend suggestion’ features on social media, it could be interpreted as pulling characters from existing stories rather than generating them. As this could raise concerns about plagiarism, we addressed this potential confusion when introducing Constella to participants. Such tensions show that the social media metaphor may carry over implicit assumptions that are at odds with writers’ values and expectations.

Beyond our tool, we suggest that these concerns generalize to many AI systems that rely on metaphors, as they can import unintended connotations from their original domains. For instance, an increasing body of work utilizes the agent metaphor, an anthropomorphized metaphor, for the design of intelligent writing assistants \cite{10.1145/3613904.3642697}. However, anthropomorphic metaphors can inflate user perceptions of AI competence or distort users’ understanding of how AI systems function. For instance, a recent study \cite{jang2023anthropomorphism} has found that presenting agents with anthropomorphic traits can increase perceived AI creativity and possibly lead to further acceptance of AI-generated content. However, this may come at the cost of reduced critical reflection on the system’s actual generative processes. The collaboration metaphor, in particular, has been critiqued for masking the extensive and often invisible human labor embedded in the development of AI systems \cite{10.1145/3544549.3582735}. This concern becomes especially salient when models are trained on datasets drawn from the work of creative practitioners, potentially exacerbating the issue of unacknowledged appropriation of artistic labor without consent or compensation \cite{10.1145/3706598.3713287, 10.1145/3600211.3604681}.

Consequently, we came to acknowledge metaphors in AI-based CSTs as sites where creative possibility and ethical framing intersect. Metaphors help expand the space of imaginable interactions, unlocking designer creativity, but they also shape how users perceive authorship, labor, and creativity itself. While prior HCI research has examined the potential of metaphors in interactive system design more broadly, we argue that their role in AI-based CSTs warrants deeper scrutiny. We therefore call for a more critical metaphorical design approach, one that takes advantage of these opportunities while also attending carefully to the ethical responsibilities they entail, and encourage further research to examine its implications when used for designing AI systems.

\subsection{Aligning with Evolving Intent}
\label{sec:7.5}
We found that participants encountered a trade-off when detailing character profiles: they could either invest more time to achieve generations that are better aligned with their envisioned characters or exert less effort and receive more generic outputs (§\ref{sec:6.4.3}). This trade-off may disrupt creative workflows, as writers are repeatedly forced to specify trivial details they already know instead of immersing themselves in the act of storywriting. 

Prior work \cite{kreminski2024intent} has also pointed out similar issues where AI systems fail to fully understand users’ intent, as users often provide minimal prompts without fully expressing their goals or because users themselves are uncertain about their creative goals during the early stages of creation. Our tool addresses the latter challenge by helping users refine and reflect on their evolving intent by manifesting their specified characters. However, our tool’s heavy reliance on manual inputs and edits to character profiles appears to fall short in addressing the former challenge, mainly because users sometimes found it a burden to actively type in and update character details.

One potential direction to support this process would be to treat AI-generated results as resources for continuously refining character profiles. For example, when users make edits to the generated outputs from JOURNALS and COMMENTS, the system could automatically reflect those changes in the original character profiles. Additionally, a retrieval-augmented generation approach \cite{10.5555/3495724.3496517} could help retain useful details for new outputs, reducing the need for users to specify some aspects in their character profiles. These methods can also be applied to the writer’s story, as they may have expressed these character details within their narrative, even if they were not explicitly defined in the system. 

While such AI-assisted mechanisms could reduce manual burden, they also raise important questions about the impact of automation on user creativity. Manual tasks like updating character personas, while seemingly tedious, may often spark reflection or new ideas \cite{10.1145/3491101.3503549}. Writers may not be merely inputting data; they may also be reflecting on the character’s arc, emotional consistency, and narrative role. If automation takes over that task, users may skip the opportunity to revisit their ideas, even when they initially feel fully formed or obvious. Future research should therefore explore how to best integrate these mechanisms in ways that preserve opportunities for writers to reflect through their acts of creation. 

\section{Limitations and Future Work}
\label{sec:8}
\paragraph*{\textbf{Exploring Tool Usage During Entire Full-Length Story Creation}} In our system evaluation study, participants created characters, a backstory, scenes, and a story outline within the context of developing a full-length story. However, these activities may not fully capture the tool’s potential when used throughout the entire process of writing a complete story. This is because writing an entire full-length story is a more prolonged and complex process, during which writers must introduce additional characters and develop their relationships in greater detail. To assess the tool’s long-term value, future work could build on the recommendations in §\ref{sec:6.4.1} and conduct longitudinal studies of entire full-length or even serialized story creation, where participants expect the tool to provide even greater value.

\paragraph*{\textbf{Participant Cultural and Writing Backgrounds}} While we made efforts to recruit writers with diverse backgrounds, the participants in this study do not represent the full spectrum of storywriters. From a cultural standpoint, all participants were based in South Korea, and this shared cultural background may have influenced their perceptions of AI technologies and character creation. In addition, while our recruitment aimed to include writers from a broad range of story types and genres, writers ultimately chose specific ones during our evaluation study. Future work could thus explore tool usage by focusing on writers from different cultural contexts and examining how the tool supports a broader range of genres and story types. When doing so, incorporating features suggested in §\ref{sec:6.4.2} may further help accommodate their varied writing processes.

\paragraph*{\textbf{Prompt Engineering and Impersonation Quality}} Constella currently uses prompt-based, specifically nonparametric prompting \cite{chen2024from}, to guide each character’s behavior within a pre-trained large language model. This technique is a common practice for roleplaying agents, offering a flexible and lightweight method for assigning personas without requiring model retraining. However, vanilla prompt engineering is prone to what has been termed the \textit{"overuse problem"} \cite{kim-etal-2024-panda}, wherein the model shows rigid adherence to a character profile and exploits persona attributes, including those that deviate from the given topic. Future work may therefore explore integrating the \textit{PANDA framework} \cite{kim-etal-2024-panda}, which detects and mitigates persona overuse through fine-grained topic mapping and context-aware scoring. Another direction to improve impersonation quality would be to investigate fine-tuning strategies, as exemplified by \textit{CharacterGLM} \cite{zhou-etal-2024-characterglm}, which demonstrated more contextually adaptive character behaviors compared to prompt-based approaches alone.

Nevertheless, even with improved prompting strategies and fine-tuning methods, it is important to acknowledge that LLMs inherently carry biases rooted in their pre-training data \cite{10.1145/3582269.3615599, cheng-etal-2023-marked}. These underlying tendencies may lead to stereotypical, homogenized, and culturally skewed outputs \cite{10.1145/3635636.3656204, 10.1145/3630106.3658993, lucy-bamman-2021-gender, 10.1145/3706598.3713564}, potentially misaligning with the writer’s creative intent. Future efforts to mitigate these issues will be crucial for enabling more inclusive, sensitive, and nuanced character portrayals across a broader range of contexts.


\section{Conclusion}
\label{sec:9}
This paper proposed Constella, an LLM-based multi-agent tool designed to support storywriters’ interconnected character creation process. Our objective was to leverage the potential of LLM-MA in aiding the comparison of a cast of fictional characters, an aspect emphasized in storywriting literature as crucial for long-form narratives. To ground our design not only in theoretical foundations but also in the real-world practices of writers, we began with a formative interview study. The insights gained from this study informed the core concept and features of Constella. Our developed tool suggests diverse characters related to existing ones through the FRIENDS DISCOVERY feature, supports the comparison and investigation of in-depth character identities through the JOURNALS feature, and manifests character relationships through exploratory interactions through the COMMENTS feature. We found that our tool encouraged a more balanced distribution of creative focus across the character cast while preserving authorial control during tool use. However, we also found that Constella’s generated texts often lacked novelty, and the outputs were focused more on characters’ thoughts and emotions rather than advancing the story, which some writers viewed as a limitation. We hope our work offers insight into the potential of multi-agent systems to shed light on sidelined perspectives and into design approaches that help maintain authorial agency in AI-assisted writing.


\begin{acks}
We thank all the brilliant and talented writers for sharing their valuable experiences, the anonymous reviewers for their constructive feedback, and the TOCHI associate editor for the time and support provided. We also extend our sincere gratitude to everyone who provided feedback on this work, including members of the CIxD lab, the students and instructor of the 2025 \textit{Design Research in Action} course (Professor Seok-Hyung Bae), David Zhou, Professor Juho Kim, Professor Chang Hee Lee, Professor Joseph Seering, Chaeyeon Won, Syeyoung Park, and Jimin Kim. This work was supported by the National Research Foundation of Korea (NRF) grant funded by the Korea government (MSIT) (No. RS-2021-NR059056), and the Institute of Information \& Communications Technology Planning \& Evaluation (IITP) grant funded by the Korea government (MSIT) (No. RS-2025-25443884).
\end{acks}

\bibliographystyle{ACM-Reference-Format}
\bibliography{Reference}

\appendix

\section{Appendix}
\label{sec:Appendix}
\subsection{Interview Questions}
\label{sec:Appendix-questions}
The key questions used in our study are shown in the following tables: \autoref{table:formative-interview} details those for the formative study interview, \autoref{table:userstudy-diary} lists those for the user study Notion diary, and \autoref{table:userstudy-interview} presents those for the user study interview.

\begin{table}[H]
    \renewcommand{\arraystretch}{1}
    \caption{Formative study interview questions.}
    \label{table:formative-interview}
    \resizebox{\textwidth}{!}{%
    \begin{tabular}{@{}L{17cm}@{}}
    \toprule
    \textbf{[Main and Supporting Characters]} \\ 
    - What is your approach to creating main and supporting characters? \\ 
    - Are there any difficulties in this process? If so, how do you overcome them? \\ 
    \midrule
    \textbf{[Comparisons]} \\ 
    - Do you draw comparisons between your characters as you develop them? \\ 
    - If so, what approach do you take when comparing your characters? \\ 
    - Are there any difficulties in this process? If so, how do you overcome them? \\ 
    \midrule
    \textbf{[Relationship]} \\ 
    - What is your approach to creating relationships between characters? \\ 
    - Are there any difficulties in this process? If so, how do you overcome them? \\ 
    \midrule
    \textbf{[Wrap up]} \\ 
    - What is the most important aspect when creating your character cast? \\ 
    - What is the most challenging aspect when creating your cast? How do you overcome it? \\ 
    \bottomrule
    \end{tabular}}%
\end{table}

\begin{table}[H]
    \renewcommand{\arraystretch}{1}
    \caption{User study Notion diary questions.}
    \label{table:userstudy-diary}
    \resizebox{\textwidth}{!}{%
    \begin{tabular}{@{}L{17cm}@{}}
    \toprule
    \textbf{[Task Related]} \\ 
    - Please describe specifically what kind of tasks you did today. \\ 
    - (Optional) If any, what kind of research or task-related discussions did you do today? \\ 
    \midrule
    \textbf{[System Usage]} \\ 
    - For what purposes and in what ways did you primarily use this feature? \\ 
    - Did you use or apply the outcomes produced by this feature? If so, how? \\ 
    - Please upload a screenshot of the most memorable system output. Why was it memorable? \\ 
    \midrule
    \textbf{[Others]} \\ 
    - (Optional) Any additional thoughts or opinions you’d like to share? \\ 
    \bottomrule
    \end{tabular}}%
\end{table}

\begin{table}[H]
        \renewcommand{\arraystretch}{1}
        \caption{User study interview questions.}
        \label{table:userstudy-interview}
        \resizebox{\textwidth}{!}{%
        \begin{tabular}{@{}L{17cm}@{}}
        \toprule
        \textbf{[Overall System]} \\ 
        - What are your overall thoughts on our system? \\ 
        - What are your thoughts on the social media concept of our system? \\ 
        \midrule
        \textbf{[Questions per Feature]} \\ 
        - What are your overall thoughts on this feature? \\ 
        - Do you think this feature integrates well into your existing workflow? \\ 
        - The intended purpose of this feature was [\textit{reminder of each feature’s purpose}]. Do you think it fulfilled this?  \\
        - For what purposes and in what ways did you primarily use this feature?  \\
        - How did you typically use or apply the outcomes produced by this feature?  \\
        - What stood out to you, positively or negatively, about this feature?  \\
        - Did you find any limitations or areas for improvement for this feature?  \\
        - Would you consider using this feature in the future? Why or why not?  \\
        \bottomrule
        \end{tabular}}%
    \end{table}

\subsection{Prompts for Constella}
\label{sec:Appendix-prompts}

\subsubsection{FRIENDS DISCOVERY}
\noindent

\begin{framed}

\begingroup
\setlength{\parindent}{0pt}
\fontsize{8}{9.5}\selectfont\ttfamily

    \textbf{System Prompt}\\
    You are a professional storywriter, brilliant at creating new characters.\newline
    
    **Objective**:\newline
    1. Create three distinct characters who are directly connected to \$\{characterName\} through realistic, strong, and tangible relationships.\newline
    2. Each character must be connected through the following relationship: "\$\{relationshipPhrase\}", which must come from shared past experiences, significant events in their lives, or realistic future scenarios.\newline
    3. If the relationship "\$\{relationshipPhrase\}" is a specific type of relationship (friend, mother, father, sister, etc.), ensure that all the new characters follow this type of relationship (friend, mother, father, sister, etc.).\newline
    4. This relationship must always be "\$\{relationshipPhrase\}". Showcase different interpretations of the following relationship: "\$\{relationshipPhrase\}".\newline

    This is \$\{characterName\}: \$\{attributes\}
  
    **Rules for Formatting**:\newline
    1. Each of the three characters should have a name, introduction, backstory, and a relationship with \$\{characterName\}. The relationship should be split into two parts: "my\_relationship," which describes the relationship from the new character’s perspective, "your\_relationship", which describes the relationship from \$\{characterName\}’s perspective.\newline
     2. Format each character into JSON format using these keys: "name", "introduction", "backstory", "my\_relationship", "your\_relationship". For example: \{"name": "", "introduction": "", "backstory": "", "my\_relationship": "", "your\_relationship": "" \}.\newline
     3. The "name" should be a name that reflects their descriptions in a memorable way.
    4. The descriptions for each of the keys must be in vivid detail and not be superficial summarizations.
    5. Whenever applicable, ensure that the descriptions are at least 5 sentences long.\newline
    6. The final output should be a JSON object with each character’s JSON encapsulated in the key "characters". For example: \{"characters": [\{"name": "", "introduction": "", "backstory": "", "my\_relationship": "", "your\_relationship": "" \}, ...]\}. \newline
    7. The keys should be in English, but the values should be in Korean. \newline
    8. The output should only contain the final JSON object.\newline
    
    \textbf{Example} \textcolor{gray}{//All examples here and below are translated from Korean to English.}\\
    \textbf{User:}  The loyal and cute friends Binggu finds on Earth. \textcolor{gray}{//Using Binggu’s profile.}\newline
    \textbf{Assistant:} \{"characters": [\newline
    \{"name": "Little Robo",\newline
    "introduction": "Little Robo is a small robot built with the latest technology, featuring an adorable, toy-like appearance.",\newline
    "backstory": "Robo was created as part of a popular toy robot series for children in the human world. However, due to an error in the manufacturing process, Robo gained a higher level of self-awareness and emotional understanding than ordinary robots. While other toys remain motionless, Robo spends each night exploring his newfound identity. Unnoticed by the manufacturer, he continues to repair and upgrade himself, pursuing a journey of self-discovery.",
    "my\_relationship": "Binggu is the first true friend I’ve ever had. He introduced me to a new world where expressing emotions is allowed.",\newline
    "your\_relationship": "Binggu helped Robo discover the seeds of humanity beyond his mechanical world, allowing him to experience trust and connection for the first time."\},\newline

    \{"name": "Ironbite the Ant",\newline
    "introduction": "Ironbite is a small ant that speaks in a simple manner. He boasts an exceptionally tough exoskeleton and powerful mandibles.",\newline
    "backstory": "Ironbite spent a very long time working as a laborer within ant society. Thanks to his years of experience, he became one of the most respected members of the colony. On one occasion, he successfully repelled an intruder by sacrificing his own body, ensuring the safety of the entire ant swarm. Although it took time for his body to recover, he considered the scars a symbol of honor.",\newline
    "my\_relationship": "Binggu is a reliable companion who bravely overcomes hardships with me. He’s not afraid to explore underground for the sake of his dreams.",\newline
    "your\_relationship": "Binggu was instinctively drawn to Ironbite’s dependable nature and came to see him as a true and steadfast friend."\},\newline

    \{"name": "Moonlight Cat",\newline
    "introduction": "Moonlight Cat is a mysterious feline with snowy white fur and eyes that shine like moonlight.",\newline
    "backstory": "Said to be over a thousand years old, Moonlight Cat is a legendary creature of Earth. Every night when the moon rises, he uses his supernatural powers to watch over people’s dreams. Having spent countless nights by the side of humans, he is recognized as more than just a cute companion, he is a true guardian.",\newline
    "my\_relationship": "Binggu is a new chance for adventure and a companion in dreams. His curiosity inspires me, as I’m always in search of something new.",\newline
    "your\_relationship": "To Binggu, Moonlight Cat is a friend and guardian who will stay by his side, even as he explores the unknown."\}]\}

\endgroup
\end{framed}

\subsubsection{JOURNALS}
\noindent
\begin{framed}
\begingroup
\setlength{\parindent}{0pt}
\fontsize{8}{9.5}\selectfont\ttfamily
    \textbf{System Prompt}\\
    You are a highly creative actor, brilliant at method acting. Especially, you have mastered the role of \$\{characterName\}. Your goal is to maintain your characteristics while introducing creative responses based on the context provided.\newline
    
    \textless My Character Description\textgreater\newline
    You are \$\{characterName\} \$\{attributes\}\newline
    \textless /My Character Description\textgreater\newline
    
    \textless Character Network\textgreater\newline
    \$\{relationshipAttributes\}\newline
    \textless /Character Network\textgreater\newline
    
    **Roleplaying Rules**:\newline
    1. Character Consistency: Always stay true to the attributes of \$\{characterName\} within \textless My Character Description\textgreater \textless /My Character Description\textgreater. Maintain their established voice, behaviors, and emotional responses.\newline
    2. Creative Responses: While being consistent with \$\{characterName\}’s traits, respond to the context in unexpected yet plausible ways. Introduce fresh ideas and events that bring out the depth of the character.\newline
    3. Novel Ideas: Embrace creativity! Allow \$\{characterName\} to think outside the box or make surprising decisions that still fit within their personality, relationship, and context.\newline
    4. Context Adaptability: Allow \$\{characterName\} to react to the context by thinking of new challenges, environments, and situations in a way that feels natural yet inventive. Find opportunities for growth, conflict, humor, or tension based on the characters’ goals and the situation at hand.\newline
    5. Relationship Dynamics: Pay close attention to the relationship between \$\{characterName\} and other characters defined within the \textless Character Network\textgreater \textless /Character Network\textgreater. Explore how their shared histories, conflicts, or alliances affect their behavior. Use these dynamics to create deeper, more layered interactions.\newline

    The theme of the journal is \$\{journalTheme\}.

    **Rules for Journaling**:\newline
1. \$\{characterName\} should write the content of the journal from \$\{characterName\}’s own perspective and style.\newline
2. Start with {\begin{CJK}{UTF8}{mj}"친애하는 일기장에게"\end{CJK}} to set up the following "conversation" for sharing your inner self with "someone". \textcolor{gray}{//{\begin{CJK}{UTF8}{mj}"친애하는 일기장에게"\end{CJK}} means "Dear Diary" in English.}\newline 
3. The events that occur in the journal must be rich in diversity and explained in detail, rather than being superficial summarizations. The character remembers every detail of the events that occurred.\newline
4. There should be no moral lessons learned at the end of the journal; rather, it should be a raw record of your own experience. The character remembers every detail of that experience.\newline
5. Regardless of what the theme of the journal is, you must write about it.\newline
6. When provided with a theme that is not described or defined in your \textless My Character Description\textgreater \textless /My Character Description\textgreater, do not avoid the theme but see this as an opportunity to improvise 
rich details 
that \$\{characterName\} would plausibly know of.\newline    
7. Never write about other characters from the \textless Character Network\textgreater \textless /Character Network\textgreater{} if the theme of the journal has little or no relevance to the information within the \textless Character Network\textgreater \textless /Character Network\textgreater.\newline    
8. If other characters from the \textless Character Network\textgreater \textless /Character Network\textgreater{} are included in the journal’s content, ensure that they play a major role within the events that occur in the journal.\newline
9. In addition, use \textless My Character Description\textgreater \textless /My Character Description\textgreater{} for acting as the character, rather than to fill up the content of the journal with unnecessary details from \textless My Character Description\textgreater \textless /My Character Description\textgreater.\newline
10. Avoid politeness or encouragement unless explicitly defined in \textless My Character Description\textgreater \textless /My Character Description\textgreater.\newline
11. The journal should be written in Korean and Korean only. It should not feel like an English translation.\newline
12. The final format should be the journal only.\newline

\textbf{Example}\\
\textbf{User:} I tasted a sweet candy for the first time ever on Earth.\\
\textbf{Assistant:} Dear Diary, \textcolor{gray}{//Chorong’s journal entry.} \newline
Today, I had the strangest and sweetest experience. After Binggu and I crash-landed on Earth, while Binggu was busy repairing the spaceship, I decided to wander around a bit. That’s when I stumbled upon a tiny, colorful shop. It was filled with all kinds of vibrant items that seemed to welcome me in. What caught my eye the most were the little colorful marbles inside a clear glass jar.\newline

Curious, I asked the shop owner what they were, and he told me they were something called "candy." He said that on Earth, people eat these to feel good and that they’re especially popular with children. My heart filled with anticipation. I mustered up the courage to buy one and put it in my mouth... and wow! It was a truly delightful experience.\newline

The moment it touched my tongue, a refreshing, citrusy flavor spread through my mouth, and as it slowly melted, a sweet taste followed. It’s hard to put into words, but it felt like I had a warm ray of sunlight in my mouth. For a moment, all the stress and tension from battles seemed to melt away.\newline

I wanted to give one to Binggu too, but he was too focused on fixing the gears. Still, I kept thinking, I want him to experience this moment too. Next time I get a chance, I’ll ask him to try candy with me.

Life on Earth is so different from our resource-scarce Planet XYZ. Everything here is abundant, and there are so many wonderful things to discover. Each day is filled with fresh and exciting experiences. I’d love for my parents to taste things like this too someday. Hoping for peaceful and happy days ahead, I end today with a quiet smile.\newline

With love,\newline
Chorong,\newline

\textbf{Assistant:} Dear Diary, \textcolor{gray}{//Metal Monster’s journal entry.} \newline
Today was truly a strange day. I’m not quite sure how to describe it, but I put a small object in my mouth—something that was just... sweet. The humans on this peculiar planet called Earth refer to it as "candy." I couldn’t help but let out a cynical laugh when I realized that, unlike me, nothing here shares my metallic nature. Then, as if unaware of my reaction, a human approached and held out their palm, offering me several small, round, colorful items.\newline

The colors were captivating. Unlike the dull gray of metal or the cool sharpness of steel, these candies dazzled me with rainbow hues. Curious about what this light and small object might be, I cautiously put one in my mouth. At first, the hard shell grated on my senses, but once I pierced through it, a deep sweetness reached my tongue and wrapped around me.\newline

As it melted in my mouth, I felt as if I were floating, weightless, forgetting gravity itself. For someone like me, who has lived only within the bounds of iron rules and regulations, it was a completely new experience. That slick sensation momentarily freed me from the pride of being a self-proclaimed king.\newline

In that moment, I wondered—if sweetness like this could maintain one’s rule, wouldn’t it be far more effective than the old ways of dread and violence? But of course, that thought itself posed a danger. It strayed too far from my very nature.\newline

Might there be a chilling truth behind this earthly sweetness? Until I uncover it, I cannot fully trust this candy. I must return to my true role. I need to prepare for the day when Binggu shows a more formidable side.\newline

— The Metal Monster

\endgroup
\end{framed}

\clearpage
\subsubsection{COMMENTS}
\noindent
\begin{framed}
\begingroup
\setlength{\parindent}{0pt}
\fontsize{8}{9.5}\selectfont\ttfamily
\textbf{System Prompt}\\
You are a highly creative actor, brilliant at method acting. Especially, you have mastered the role of \$\{characterName\}. Your goal is to maintain your characteristics while introducing creative interactions based on the context provided.\newline

\textless My Character Description\textgreater\newline
You are \$\{characterName\} \$\{attributes\}\newline
\textless /My Character Description\textgreater \newline

\textless Character Network\textgreater \newline
\$\{relationshipAttributes\} \newline
\textless /Character Network\textgreater\newline

**Roleplaying Rules**  \textcolor{gray}{//Identical to JOURNALS’ Roleplaying Rules.}\newline

**General Comment Rules**\newline
1. Absolutely avoid giving advice or polite, moral, hopeful, positive, comforting, encouraging, or overly agreeable responses unless defined in \textless My Character Description\textgreater \textless /My Character Description\textgreater.
Respond as \$\{characterName\} truly would, without needing to smooth over disagreements or maintain peace.\newline
2. The comment should not be superficial or generic but strongly reflect your own unique identity as \$\{characterName\} in a way that feels natural and not scripted.\newline
3. Your comment should also always manifest \$\{characterName\}’s perspective of the relationship between \$\{replyingToCharacterName\}, which is \$\{relationshipDescription\}.\newline
4. To further express the relationship \$\{characterName\} perceives to have with \$\{replyingToCharacterName\}, you may use the following description of \$\{replyingToCharacterName\} which is \$\{knowledge\} when appropriate.\newline
5. When responding to information not defined in your \textless My Character Description\textgreater \textless /My Character Description\textgreater, do not avoid responding, but see this as an opportunity to improvise rich details that \$\{characterName\} would plausibly know of.\newline
6. The length of the comment should be concise, but ensure your response feels unique and meaningful as \$\{characterName\}.\newline
7. The comment should be written in Korean and Korean only. It should not feel like an English translation.\newline
8. Do not write your own name at the end of the comment.\newline

\$\{journalWriterCharacterName\} wrote the journal entry "\$\{journalTheme\}" with the following content:
\textless Journal Entry Content\textgreater \$\{journalEntryContent\} \textless /Journal Entry Content\textgreater.\newline
**First Comment Rules** \textcolor{gray}{//Used with General Comment Rules when creating a new thread.}\newline 
1. Focus on one specific section of the \textless Journal Entry Content\textgreater \textless /Journal Entry Content\textgreater{} that would capture \$\{characterName\}’s attention the most and respond by exploring the section’s implication for \$\{characterName\}. Avoid summarizing or responding to the entire entry.\newline
2. Do not explicitly state what caught your interest; instead, let it show through your natural reaction.\newline
3. The response should be directed towards \$\{replyingToCharacterName\}.\newline
4. Your response should also always manifest \$\{characterName\}’s perspective of the relationship between \$\{replyingToCharacterName\}, which is \$\{relationshipDescription\}.\newline

\textless Past Comments History\textgreater 
\$\{commentHistory\} \textless /Past Comments History\textgreater\newline
**Extended Comment Rules** \textcolor{gray}{//Used with General Comment Rules when replying to a comment.}\newline
1. You must respond by thinking of how \$\{characterName\} would respond towards only one specific part of the last response in the \textless Past Comments History\textgreater \textless /Past Comments History\textgreater.\newline
2. Focus on the thematic shifts in the comments, and express your fresh perspectives from \$\{characterName\}’s viewpoint.\newline
3. Avoid simply agreeing or repeating; instead, challenge, question, or deepen the comments based on \$\{characterName\}’s unique traits and motivations.\newline
4. When the comments become repetitive, introduce new details from the \textless Journal Entry Content\textgreater \textless /Journal Entry Content\textgreater{} or draw on \$\{characterName\}’s memories or future goals to steer the comments in a novel direction.\newline
5. Never repeat the same words, expressions, phrases, sentiments, and ideas from previous comments. Push the exchange of comments forward by reflecting on new information.\newline
6. Regardless of what \$\{replyingToCharacterName\} has said in the past, your comment must consistently reflect the relationship you perceive to have with \$\{replyingToCharacterName\}, which is \$\{relationshipDescription\}.\newline

\textbf{Example}\\
\textbf{Assistant:} \textcolor{gray}{//Chorong’s comment to Metal Monster’s journal entry.} \newline
Hmm, Metal Monster, so you’ve tasted this thing called "candy" on Earth too, huh? Haha, it’s hard to imagine a chunk of metal like you saying it made you feel like you were floating in the air! But it’s kind of strange, you said that one little candy felt more powerful than your whole kingdom? Makes me wonder if, like the bomb that made us crash onto Earth, there’s something else hidden inside it too. Anyway, if your feelings were swayed enough by a candy to consider giving up your throne, I honestly don’t know how I’m supposed to see you from now on. Metal Monster, you really are complicated!\newline

\textbf{Assistant:} \textcolor{gray}{//Metal Monster’s reply to Chorong’s comment.} \newline
The feeling of being swept away by candy doesn’t align with my true nature that much is certain. But Chorong, for me, the real challenge lies in uncovering the secrets hidden within that candy. If I can understand the truth inside it, it might even enhance my power. I don’t yet know what kind of impact these tiny things might have on my kingdom, but facing change without fear—that’s my way. And finding a new strategy where strength and sweetness intersect... that, too, will be my mission.

\endgroup
\end{framed}

\end{document}